\pdfoutput=1
\documentclass[8.5pt,twocolumn]{article}
\usepackage[super,sort&compress,comma]{natbib} 
\usepackage{mhchem}
\usepackage{times,mathptmx}
\usepackage{sectsty}
\usepackage{balance} 
\usepackage{color} 

\usepackage{geometry}
\usepackage{amsmath}
\usepackage{amssymb}
\usepackage{float}

\newcommand{\rev}[1]{{\textcolor{black}{#1}}}

\usepackage{graphicx} 
\usepackage{lastpage}
\usepackage[format=plain,justification=raggedright,singlelinecheck=false,font=small,labelfont=bf,labelsep=space]{caption} 
\usepackage{fancyhdr}

\begin{document}


\makeatletter 
\def\subsubsection{\@startsection{subsubsection}{3}{10pt}{-1.25ex plus -1ex minus -.1ex}{0ex plus 0ex}{\normalsize\bf}} 
\def\paragraph{\@startsection{paragraph}{4}{10pt}{-1.25ex plus -1ex minus -.1ex}{0ex plus 0ex}{\normalsize\textit}} 
\renewcommand\@biblabel[1]{#1}            
\renewcommand\@makefntext[1]%
{\noindent\makebox[0pt][r]{\@thefnmark\,}#1}
\makeatother 
\renewcommand{\figurename}{\small{Fig.}~}
\sectionfont{\large}
\subsectionfont{\normalsize} 

\setlength{\columnsep}{6.5mm}
\setlength\bibsep{1pt}

\twocolumn[
  \begin{@twocolumnfalse}
\noindent\LARGE{\textbf{\rev{Microfluidic Flow of Cholesteric Liquid Crystals}}}
\vspace{0.6cm}

\noindent\large{\textbf{Oliver Wiese \textit{$^{a}$}, Davide Marenduzzo \textit{$^{a}$}, Oliver Henrich \textit{$^{a,b \;\ast}$}}}\vspace{0.5cm}
\center{\small\textit{$^{a}$~SUPA, School of Physics and Astronomy, University of Edinburgh, JCMB, King's Buildings, Edinburgh EH9 3FD, UK\\
$^{b}$~EPCC, School of Physics and Astronomy, University of Edinburgh, JCMB, King's Buildings, Edinburgh EH9 3FD, UK\\
$\ast$~E-mail: ohenrich@epcc.ed.ac.uk}}



\vspace{0.6cm}

\noindent \normalsize{
We explore the rheology and flow-induced morphological changes of cholesteric liquid crystal patterns subject to Poiseuille flow within a slab geometry, and under different anchoring conditions at the wall. Our focus is particularly on the behaviour of ``Cholesteric Fingers of the first kind'' and of Blue Phase II. Depending on the applied pressure gradient, we observe a number of dynamic regimes with different rheological properties. Our results provide the first insight into the flow response of cholesteric phases with fully two- or three-dimensional director field patterns and \rev{normal and planar degenerate anchoring conditions as commonly realised in experiments. They are also of high relevance for a fundamental understanding of complex liquid crystals in confinement and an important step towards future microfluidic applications that are based on cholesteric liquid crystals.}
}
\vspace{0.5cm}
 \end{@twocolumnfalse}
  ]





\section{Introduction}

Cholesterics, or chiral nematics, are liquid crystals in which the local director field, representing the average direction of orientational order, shows spontaneous twist in thermodynamic equilibrium~\cite{deGennes,Chandrasekhar}. The simplest case is the standard cholesteric phase where the director field precesses around a single helical axis of fixed orientation. The associated director field pattern, which only varies along one dimension, is also called the ``Grandjean texture''. 

Sandwiching a cholesteric helix between two flat walls where the director field is anchored normally (or ``homeotropically'') creates frustration if the helix of the cholesteric axis is parallel to the walls. This frustration is resolved by the creation of more complex, two-dimensional director field patterns which are known as ``Cholesteric Fingers'' ~\cite{Oswald:2000} (CFs). Fully three-dimensional structures are also possible, and are encountered for highly chiral systems, for which the preferred configuration close to the isotropic boundary features twist around two perpendicular axes, as opposed to just one axis as in the regular cholesteric state. The corresponding conformation is referred to as ``double-twist cylinder'' (DTC). As it is topologically impossible to cover three-dimensional space continuously with double-twist cylinders, defects arise by necessity. The resulting disclination lines, at which the nematic director field is undefined, organise into a variety of three-dimensional lattices, giving rise to the so-called cubic Blue Phases (BPs)~\cite{Grebel:1984,Wright:1989}. There are two experimentally observed cubic Blue Phases, BPI and BPII. A third, BPIII, is thought to be amorphous ~\cite{Henrich:2011a}. In the last decade BPs have moved from little more than an academic curiosity into the forefront of liquid crystal device technology after the discovery that they can be stabilised over a much wider temperature range than previously thought possible~\cite{Kikuchi:2002,Coles:2005,Kitzerow:2006}.

\rev{Even simple cholesterics are known to show strong non-Newtonian flow behaviour that depends sensitively on the applied boundary conditions, and is due to interplay of elastic forces and order-flow coupling. When probed along the direction of the helical axis (also know as permeative mode) cholesterics can have very large apparent viscosities at small flow velocities, and typically display strong shear thinning at larger flow rates ~\cite{experimentpermeation1,Helfrich:1969,experimentpermeation2}. When the helical axis is oriented perpendicular to the flow direction (vorticity mode) an uncoiling transition occurs at a critical flow rate ~\cite{Rey:1996a,Rey:1996b}. The flow of two- or three-dimensional director field patterns such as the above mentioned is even more intricate and is far from being understood. Unsurprisingly, the presence of complex boundaries or interfaces adds another level of complexity to the problem, and may be the reason why the microfluidic flow of liquid crystals has been left vastly unexplored since the emergence of this field in the early 1990s.}

\rev{The last few years have seen a rapidly increasing interest in the flow of passive and active nematic liquid crystals in microfluidic confinement ~\cite{Sengupta:2014,Thampi:2015}. Anchoring conditions ~\cite{Batista:2015} and the defect topology ~\cite{Agha:2016} have been shown to play a crucial role for the emerging flow profile in the microchannel. Their behaviour at liquid-liquid interfaces ~\cite{Silva:2015,Majumdar:2016} and in liquid crystal emulsions ~\cite{Sivakumar:2009} have also been studied. Another promising route focusses on nonlinear electrophoretic and electrohydrodynamic aspects and transport properties of suspended nanoparticles ~\cite{Lavrentovich:2014, Sasaki:2015}. An intriguing application is the combination of liquid crystal microfluidics and optics ~\cite{Cuennet:2013,Wee:2016} for advanced displays, tuneable filters and light modulation. In this respect, microfluidic applications based on cholesteric liquid crystals with their intrinsic ability to order in 1D, 2D and 3D superstructures offer exciting new avenues to a completely new range of optofluidic applications like lasers and optical sensors ~\cite{Bisoyi:2014}.}

\rev{Our aim in this work is to study the rheology of Cholesteric Fingers and Blue Phases subjected to normal and planar degenerate anchoring of the director field at the boundary walls. This is a first step towards understanding the flow behaviour of {\it complex} liquid crystals in {\it simple} geometries. Previous research into microfluidic aspects of liquid crystals has been mostly devoted to studying {\it simple} liquid crystals (such as nematics) in {\it complex} geometries}. Importantly, while there have been a number of computer simulation studies of the rheology of \rev{confined} cholesteric liquid crystals and Blue Phases, all of these have \rev{considered either free boundary conditions where the director field can freely rotate without energetic penalties, or pinned boundary conditions where the cholesteric pattern is permanently fixed at the walls.} Pinning may provide a suitable approximation of a situation where there are impurities~\cite{AdrianoSoftMatter}. However, a more natural scenario, which is simpler and \rev{can be realised more easily in the lab, is anchoring of the director field either normal or tangential to the walls}. This can be enforced respectively either by using a surfactant or by mechanical rubbing. In our work we consider these \rev{experimentally} more realistic boundary conditions, and ask how these affect the rheological behaviour of cholesteric liquid crystals. This is an important step \rev{forward to understand the rheological and hydrodynamic aspects of these systems, and to allow a comparison between theoretical predictions and existing or future experiments and emerging applications}~\cite{Tiribocchi:2014}. At the same time, from a theoretical viewpoint, anchoring in for instance Blue Phases is of interest because it leads to frustration of the bulk topology, which in turn results in the creation of exotic defect networks. These are stable only in thin films due to the anchoring at the boundaries ~\cite{Fukuda:2010a, Fukuda:2010b, Ravnik:2011b} and can be further manipulated by means of dielectric or flexoelectric fields ~\cite{AdrianoSoftMatter,AdrianoFlexo}. 

Our paper is organised as follows. In Section \ref{methods}, we describe the methods we use and review the hydrodynamic equations of motion which we solve. In Section \ref{results}, we report our numerical results, first for our studies of Cholesteric Fingers (Section \ref{cf}), followed by those of Blue Phase II (Section \ref{bp-norm} with normal anchoring, and Section \ref{bp-plan} with planar degenerate anchoring). Finally, Section \ref{conclusions} contains our conclusions.

\section{Model and Methods}\label{methods}

Our approach is based on the well-established Beris-Edwards model for hydrodynamics of cholesteric liquid crystals ~\cite{Beris:1994}, which describes the ordered state in terms of a traceless, symmetric tensor order parameter ${\mathbf Q}({\mathbf r})$. 
In the uniaxial approximation, the order parameter is given by $Q_{\alpha \beta}= q_s ( \hat{n}_\alpha \hat{n}_\beta - \frac{1}{3}\; \delta_{\alpha\beta})$ with $\hat{{\mathbf n}}$ the director field and $q_s$ the amplitude of nematic order. More generally, the largest eigenvalue of ${\mathbf Q}$, $0\le q_s\le\frac{2}{3}$ characterises the local degree of orientational order.
The thermodynamic properties of the liquid crystal are determined by a Landau-deGennes free energy ${\cal F}$, whose density $f$ consists of a bulk contribution $f_b$ and a gradient part $f_g$, as follows,
\begin{eqnarray}
f_b&=&\frac{A_0}{2}\left(1-\frac{\gamma}{3}\right) Q_{\alpha \beta}^2\nonumber\\
&-&\frac{A_0 \gamma}{3}Q_{\alpha \beta} Q_{\beta \gamma} Q_{\gamma \alpha}+\frac{A_0 \gamma}{4}(Q_{\alpha \beta}^2)^2,\nonumber\\
f_g&=&\frac{K}{2}(\varepsilon_{\alpha\gamma\delta} \partial_\gamma Q_{\delta\beta}+2 q_0 Q_{\alpha \beta})^2+\frac{K}{2}(\partial_\beta Q_{\alpha \beta})^2.\label{FE}
\end{eqnarray}
The first term contains a bulk-free energy constant $A_0$ and the temperature-related parameter $\gamma$ which controls the magnitude of order.
The second part quantifies the cost of elastic distortions, which is proportional to the elastic constant $K$;
we work for simplicity in the one-elastic constant approximation~\cite{deGennes}. The wavevector $q_0$ is equal to $2\pi/p_0$, where $p_0$ is the cholesteric pitch, 
\rev{the length scale over which the director field rotates a full turn around the axis of the cholesteric helix}.

The actual periodicity of the BP structure, $p$, does not need to be equal to $p_0$.
Indeed, the ``redshift'' $r=p/p_0$ is adjusted during the equilibration phase of the 
simulation, to optimise the free energy density
before shearing begins. This is done by following the procedure
previously described in~\cite{Alexander:2006}.

Anchoring conditions are imposed by adding an extra term to the Landau-deGennes free energy functional at the boundaries. For homeotropic or normal anchoring, 
where the director field prefers to be oriented normal to the surfaces, this is achieved through 
\begin{equation}
\label{normal_ac}
f_s=\frac{W}{2}(Q_{\alpha \beta} - Q_{\alpha \beta}^0)^2
\end{equation}
with $Q_{\alpha \beta}^0$ as the preferred configuration of the order 
parameter tensor at the surface.
For planar degenerate anchoring conditions ~\cite{Fournier:2005} a fourth-order term has to be added:
\begin{equation}
\label{planar_ac}
f_s=\frac{W_1}{2}(\tilde{Q}_{\alpha\beta}-\tilde{Q}_{\alpha\beta}^\perp)^2+\frac{W_2}{2}(\tilde{Q}_{\alpha\beta}\tilde{Q}_{\alpha\beta}-S_0^2)^2.
\end{equation}
$S_0$ is a fixed surface amplitude. The two tensor $\tilde{Q}$ and $\tilde{Q}^\perp$ are defined as 
\begin{eqnarray}
\tilde{Q}_{\alpha\beta}&=&Q_{\alpha\beta}+\frac{1}{3} \delta_{\alpha \beta} S_0\\
\tilde{Q}_{\alpha\beta}^\perp&=&P_{\alpha\mu}Q_{\mu\nu}P_{\nu\beta}
\end{eqnarray}
with $P_{\alpha\beta}=\delta_{\alpha\beta}-n_\alpha n_\beta$ as projector onto the surface defined by its unit normal $n_\alpha$. 

A thermodynamic state is specified by two dimensionless quantities: the reduced temperature 
\begin{equation}
\tau=\frac{27(1-\gamma/3)}{\gamma},
\end{equation}
which vanishes at the spinodal point of a nematic ($q_0=0$), and the reduced chirality 
\begin{equation}
\kappa=\sqrt{\frac{108 K q_0^2}{A_0 \gamma}},
\end{equation}
which measures the ratio of gradient to bulk free energy.

The dynamical evolution of the order parameter is given by the equation 
\begin{equation}
\left(\partial_t+ v_\alpha \partial_\alpha \right){\mathbf Q} - {\mathbf S}({\mathbf W},{\mathbf Q}) = \Gamma {\mathbf H}.
\label{op-eom}
\end{equation}
The first term on the left hand side of Eq.\ref{op-eom} is a material derivative, which describes the rate of change of a quantity advected by the flow. 
The second term accounts for the rate of change due to local velocity gradients $W_{\alpha \beta}=\partial_\beta v_\alpha$, and is explicitly given by~\cite{Beris:1994}
\begin{eqnarray}
{\mathbf S}({\mathbf W}, {\mathbf Q}) &=& (\xi {\mathbf A} + {\boldsymbol \Omega})({\mathbf Q}+\frac{\mathbf I}{3})\nonumber\\
& &\hspace*{-1.5cm}+ ({\mathbf Q}+\frac{\mathbf I}{3})(\xi {\mathbf A}  - {\boldsymbol \Omega})-2 \xi ({\mathbf Q}+\frac{\mathbf I}{3})
\mathrm{Tr}({\mathbf Q \mathbf W}),
\label{sw}
\end{eqnarray}
where $\mathrm{Tr}$ denotes the tensorial trace, while ${\mathbf A}=({\mathbf W}+{\mathbf W}^T)/2$ and ${\boldsymbol \Omega}=({\mathbf W}-{\mathbf W}^T)/2$ are the symmetric and antisymmetric part of the velocity gradient, respectively. $\xi$ is a constant depending on the molecular details of the liquid crystal. Flow alignment occurs if $\xi \cos{2\theta}=(3q_s)/(2+q_s)$ has a real solution for $\theta$, the so-called Leslie angle. We select this case by setting $\xi=0.7$ in our simulations. ${\mathbf H}$ is the molecular field, which is a functional derivative of $\cal F$ that respects the tracelessness of $\mathbf Q$:
\begin{equation}
{\bf H}=-\frac{\delta {\cal F}}{\delta {\bf Q}}+\frac{\bf I}{3}\,
\mathrm{Tr} \left(\frac{\delta {\cal F}}{\delta {\bf Q}}\right).
\label{molfield}
\end{equation}
The rotational diffusion constant $\Gamma$ in Eq.\ref{op-eom} is proportional to the inverse of the rotational viscosity $\gamma_1=2 q_s^2/\Gamma$~\cite{deGennes}.

The time evolution of the fluid density and velocity are respectively governed by the continuity equation $\partial_t \rho = -\partial_\alpha(\rho v_\alpha)$, and the following Navier-Stokes equation:
\begin{eqnarray}
\partial_t v_\alpha +\rho \,v_\beta \partial_\beta v_\alpha
&=& \partial_\beta \Pi_{\alpha \beta}+ \nu_0\, \partial_\beta [ \partial_\alpha v_\beta +\; \partial_\beta v_\alpha].
\label{NSE}
\end{eqnarray}
This emerges from the Chapman-Enskog expansion of the lattice Boltzmann (LB) equations ~\cite{Succi,Guo} that we solve numerically. A further term $\nu_0(1+3\frac{\partial P_0}{\partial\rho} )\partial_\mu v_\mu \delta_{\alpha \beta}$ that formally appears in this expansion is negligible under the slow flows considered here for which the fluid motion is almost incompressible~\cite{Denniston:2001}. In Eq.\ref{NSE}, $\nu_0$ is an isotropic background viscosity which is set to $\nu_0=5/3$ in LB units (see discussion below). The thermodynamic stress tensor reads explicitly
\begin{eqnarray}
\Pi_{\alpha \beta}&=&P_0 \delta_{\alpha\beta}
-\xi H_{\alpha \gamma}\left(Q_{\gamma \beta} +\frac{1}{3} \delta_{\gamma \beta}\right)\nonumber\\
&-&\xi \left(Q_{\alpha \gamma} +\frac{1}{3} \delta_{\alpha \gamma}\right) H_{\gamma \beta} + Q_{\alpha \gamma}H_{\gamma \beta}-H_{\alpha \gamma} Q_{\gamma \beta} \nonumber\\
&+&2 \xi  \left(Q_{\alpha \beta} +\frac{1}{3} \delta_{\alpha \beta}\right) Q_{\gamma \nu} H_{\gamma \nu}
- \partial_\alpha Q_{\gamma \nu} \frac{\delta{\cal F}}{\delta \partial_{\beta} Q_{\gamma \nu}}\nonumber\\
\label{Pi}
\end{eqnarray}
and is responsible for strong non-Newtonian flow effects. In the isotropic state ${\bf Q}\equiv 0$ and Eq.\ref{Pi} reduces to the scalar pressure as would appear in Eq.\ref{NSE} for a Newtonian fluid. 

We next define a dimensionless number that describes the deformation of the director field under flow. This so-called Ericksen number is given by 
\begin{equation}
Er=\frac{\nu_0 V l}{K}
\end{equation}
with $\nu_0$ and $K$ defined previously, and $V$ and $l$ a typical velocity and length scale. In the present work $l=p_0/2=\pi/q_0$ was used as this is the approximate size of the BP unit cell. Likewise, $V$ was taken to be the velocity difference across one unit cell, i.e. $V=\dot{\gamma}\, l$. 
The apparent viscosity $\nu_{app} = \Phi / \Phi_0$ is found as ratio between the volumetric flow rate 
\begin{equation}
\label{volumetric_flow_rate}
\Phi = \int_0^{L_x}\int_0^{L_y} v_z(x)\; dx \,dy \\
\end{equation}
through a plane perpendicular to the flow or z-direction and the flow rate $\Phi_0$ of a Newtonian fluid at the same pressure gradient. The volumetric flow rate $\Phi_0$ of a Newtonian fluid with dynamic viscosity $\nu_0$ through a gap $L_x$ driven by a pressure difference $\Delta p = f \, L_z$ in plane Poiseuille flow is 
\begin{eqnarray}
\label{volumetric_flow_rate_0}
\Phi_0 &=& \int_0^{L_x} \int_0^{L_y} \frac{L_x^2}{2\nu_0}\frac{\Delta p}{L_z} \left(\frac{x}{L_x} - \left(\frac{x}{L_x}\right)^2\right) dx \,dy\\
&=& \frac{L_x^3 \,L_y \,f}{12\,\nu_0}.
\end{eqnarray}
Here, the parameter $f$ plays the role of a forcing parameter which is equivalent to a homogenous pressure gradient between inlet and outlet.

The system of coupled partial differential equations, Eqs.\ref{op-eom} and\ref{NSE}, is solved by means of a hybrid method~\cite{Marenduzzo:2007} which uses a combination of lattice Boltzmann and finite difference schemes.
(This is in contrast with some earlier methods using solely LB~\cite{Denniston:2001, Denniston:2004}.)
The Navier-Stokes equation is solved via the lattice Boltzmann approach, using a standard three-dimensional model with 19 discrete velocities (D3Q19).
A regular lattice with spacing $\Delta x = \Delta y = \Delta z = 1$ is used and the time step is $\Delta t = 1$ in lattice units.
Coupling to the thermodynamic sector is via a local body force computed as the divergence of the thermodynamic stress Eq.\ref{Pi}; the resulting velocity field is used in the computation
of the time evolution of $\mathbf{Q}$ via a standard finite difference method using the same grid and the same time step as the LB.

\begin{figure}[htpb]
\begin{center}
\includegraphics[width=0.45\textwidth]{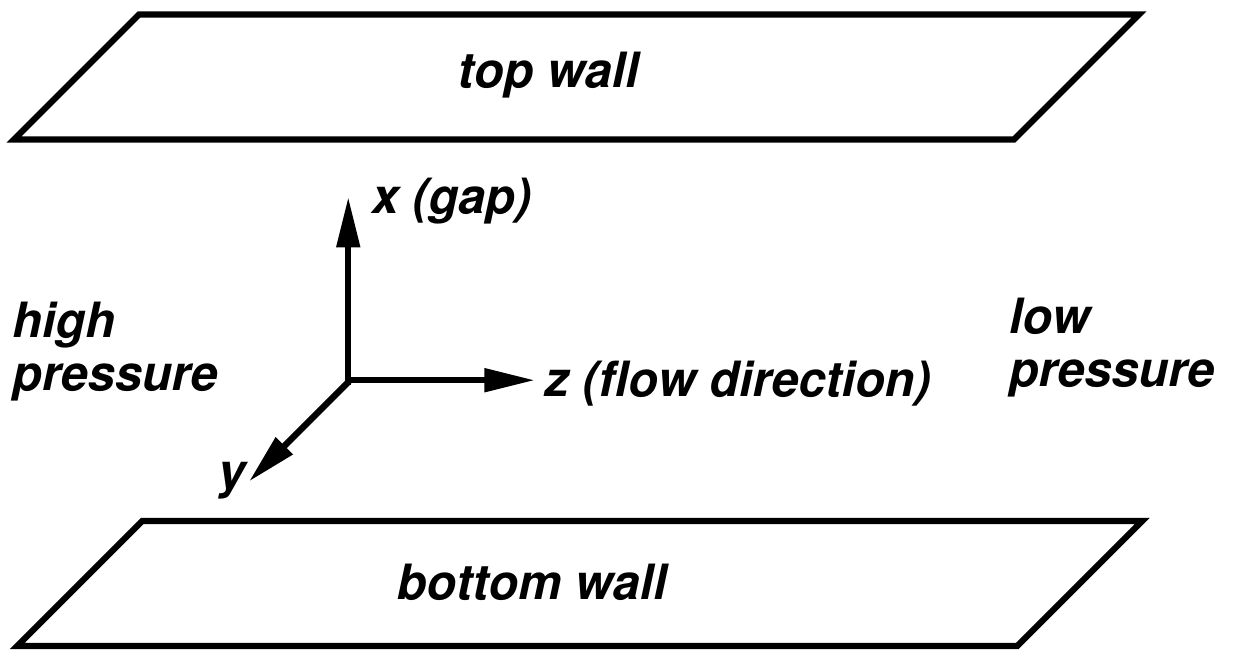}
\caption{\label{schematic} \rev{Schematic picture of the simulation geometry. At the y- and z-boundaries we applied periodic boundary conditions, whereas the x-boundaries are modelled as solid walls with no-slip boundary conditions for the fluid flow and normal or planar degenerate anchoring conditions for the liquid crystal.}}
\end{center}
\end{figure}

\rev{The system was set up with solid walls with no-slip boundary conditions at the top and bottom x-boundary and with periodic boundary conditions along the y- and z-direction: hence the geometry is that of an infinitely extended channel, gap or slit pore. The pressure gradient $f$ between inlet and outlet was applied as a homogeneous body force acting at every point within the system so that the total pressure difference between inlet and outlet becomes $\Delta p=f\,L_z$. Note that in the present work the flow is along the z-direction, whereas velocity gradients are along the x-direction. A schematic picture of our simulation geometry is given in Fig. \ref{schematic}.}

The timestep $\Delta t$ and lattice spacing $\Delta x$ in lattice Boltzmann units (LBU) can be mapped approximately to $\Delta t \sim 1 {\rm ns}$ and $\Delta x \sim 10 {\rm nm}$ in SI units, respectively. The LB unit of stress or pressure is equal to about $10^8 {\rm Pa}$. \rev{A pitch length $p_0$, in the range of $200 {\rm nm}$ to $1{\rm \mu m}$ in SI units is resolved by 32 (BPII) or 64 (CF1) lattice sites. This means the gap along the x-direction between the walls corresponds to approximately $L_x = 0.5 - 2 {\rm \mu m}$ depending on the pitch length. It is worth mentioning that, owing to the top-down nature of our model, these values for time, length unit and pressure scale put the elastic constant $K=0.04$ in LBU on the larger side of experimentally reported values. Our results, however, did not show sensitivity towards the specific value of the elastic constant as long as they were chosen to be sufficiently small for the numerical procedure.} Further details about the conversion from LBU to SI units can be found in~\cite{Henrich:2010b,Henrich:2011a}.

\section{Results and Discussion}\label{results}

\subsection{Cholesteric Fingers}\label{cf}

\begin{figure}[htpb]
\begin{center}
\includegraphics[width=0.45\textwidth]{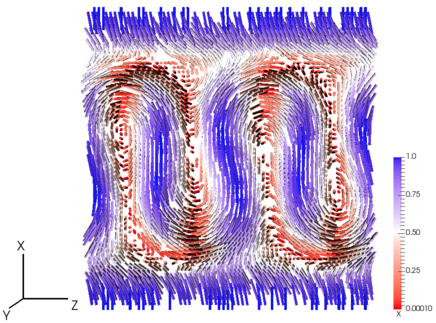}
\caption{\label{CF1_init} Director field of a quiescent Cholesteric Finger of the first kind (CF1) after equilibration. \rev{The colour code highlights the x-component of the director field, which is the component perpendicular to the walls. The colour ranges from blue (director fully along x) over white to red (director in y-z plane)}.}
\end{center}
\end{figure}

A Cholesteric Finger of the first kind (CF1, see Fig.\ref{CF1_init}) is an effectively two-dimensional conformation, which is translationally invariant along the third dimension, that defines their so-called finger axis. Making use of this symmetry, we studied 2D slices along the y-direction of size $L_x \times L_y \times L_z = 64 \times 1 \times 64$. \rev{The colour code gives the x-component of the director field perpendicular to the walls; the colour ranges from blue (fully perpendicular) over light blue, white, orange to red (director in the y-z plane).} At the walls, situated at the top and bottom boundary along the x-direction, we applied strong homeotropic anchoring conditions with an anchoring strength of $W/K=10$, corresponding to strong anchoring (which is required to stabilise CF1s thermodynamically). Note that $W \Delta x/K$ would define a dimensionless parameter. We omit the lattice constant from this definition as in lattice Boltzmann $\Delta x =1$ in LBU.

Fig.\ref{CF1_init} shows the initial state after equilibration phase, before the pressure gradient is switched on. The CF1 structure is symmetric with respect to a point in the middle of the channel where the director field escapes into the third dimension. Note that Cholesteric Fingers of the second or third kind exhibit different symmetries~\cite{Oswald:2000}, and are not considered here. 

\begin{figure}[htpb]
\begin{center}
\includegraphics[width=0.45\textwidth]{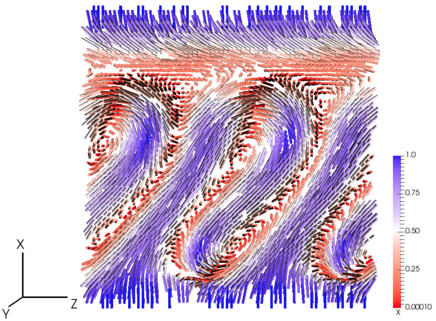}
\caption{\label{CF1_tiltedFinger} Director field of a drifting CF1, for a pressure gradient of $f=10^{-5}$. The colour code highlights the x-component of the director field, which is the component perpendicular to the walls from \rev{blue (fully perpendicular) over white to red (director in y-z plane)}. The flow occurs along the horizontal direction from left to right.}
\end{center}
\end{figure}

When a pressure gradient is applied between inlet and outlet, the CF1 structure adopts different morphologies depending on the strength of the forcing. 
At the lowest applied pressure gradients $f\lesssim 5\times 10^{-7}$, we found that the quiescent CF1 conformation of Fig. \ref{CF1_init} got slightly distorted during the onset of the flow, but then ceased to move visibly. While the entire conformation had a quasi-static appearance, a finite mass flow through the channel could be measured, providing evidence for an almost purely permeative flow, where the liquid crystal can only flow by simultaneously rotating its director field so as to leave the overall director field pattern unmodified over time.

\begin{figure}[htpb]
\begin{center}
\includegraphics[width=0.45\textwidth]{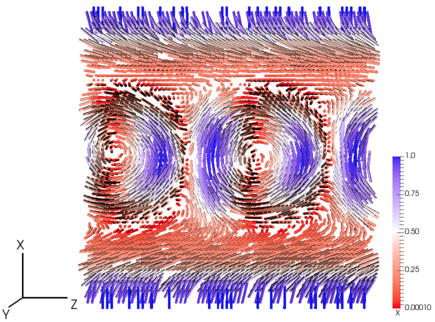}
\caption{\label{CF1_driftBubble} Director field of a CF1 under flow, for a pressure gradient of $f=5\times 10^{-5}$, resulting in bubble-like conformations. The colour code highlights the x-component of the director field, which is the component perpendicular to the walls. \rev{The colour ranges from blue (director fully along the x-direction) over white to red (director in y-z plane)}. The flow occurs along the horizontal direction from left to right.}
\end{center}
\end{figure}

\begin{figure}[htpb]
\begin{center}
\includegraphics[width=0.45\textwidth]{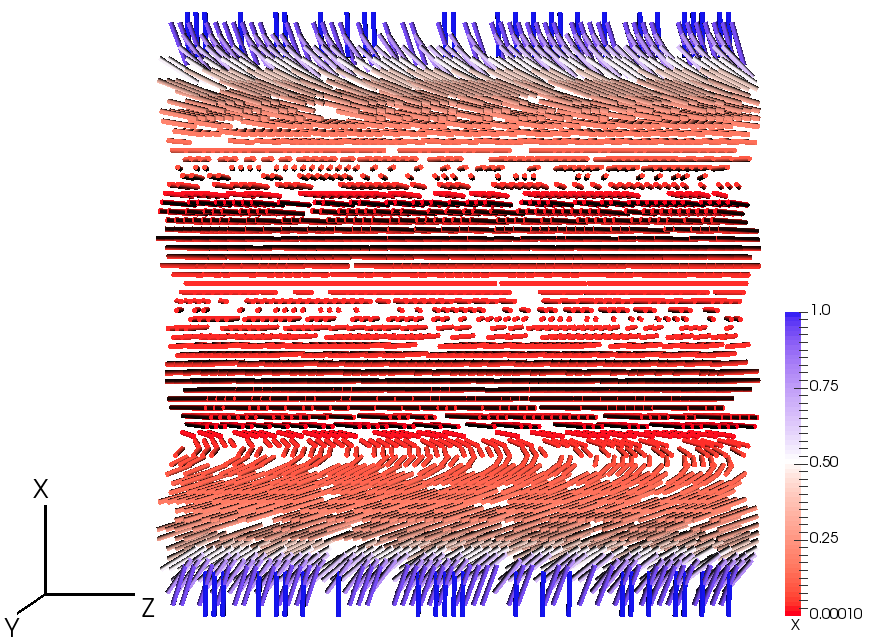}
\caption{\label{CF1_bands} The CF1 turns into a simpler, Grandjean-like conformation for a large pressure gradient, of $f= 10^{-4}$. The colour code highlights the x-component of the director field, which is the component perpendicular to the walls from \rev{blue (fully perpendicular) over white to red (director in y-z plane)}. The flow occurs along the horizontal direction from left to right.}
\end{center}
\end{figure}

At slightly larger pressure gradients we observed various dynamical regimes. Figures \ref{CF1_tiltedFinger}, \ref{CF1_driftBubble} and \ref{CF1_bands} show director profiles in the steady state for different pressure gradients.  The pressure gradient is oriented along the horizontal axis so that the flow direction is from left to right.
For pressure gradients larger than the above stated but below $f<2.5\times 10^{-5}$ we observed tilting, drifting and stretching of the finger pattern. A snapshot of the conformation is shown in Fig. \ref{CF1_tiltedFinger} for a typical value of $f=10^{-5}$. \rev{Interestingly, this flowing CF1 structure features an asymmetric director profile pattern with respect to the centre line of the channel. At the bottom the finger pattern appears connected to the wall (this is apparent from following the blue layers), while at the top the cholesteric layers detach from the wall and form a region with an orientation perpendicular to the wall. This asymmetric detachment of the finger pattern results as a consequence of the symmetry of the director field with respect to the finger axis of the CF1 (the regions enclosed by the undulating pattern at the centre of the channel). In other words, the two sides of the channel are not related through a mirror symmetry. Hence, the order-flow coupling on the two sides is different, and this leads to the one-sided detachment.}

At larger pressure gradients $2.5\times 10^{-5}\le f \le 5\times 10^{-5}$ the stretched and dilated finger pattern exists only transiently before another steady state is reached. This state is depicted in Fig. \ref{CF1_driftBubble} and consists of bubble-like domains at the centre of the channel, roughly the size of the pitch length. These domains are now on both sides separated from the walls by regions where the director field is oriented primarily parallel to the walls. Interestingly, this dynamical regime is observed only over a narrow range of pressure gradients. 

\begin{figure}[htpb]
\begin{center}
\includegraphics[width=0.495\textwidth]{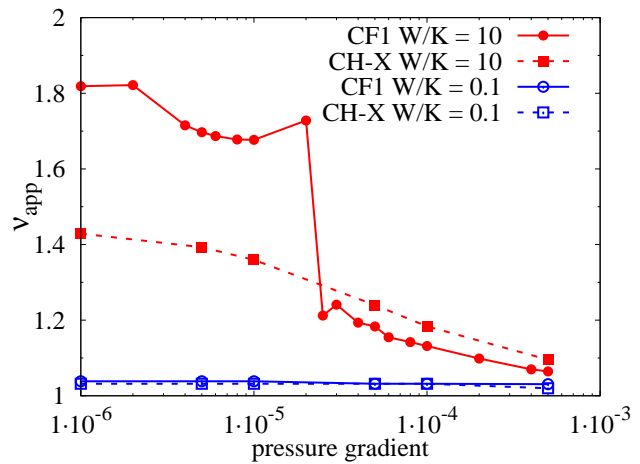}
\caption{\label{CF1_viscosity} Apparent viscosity vs pressure gradient for the CF1 \rev{in the weak and strong homeotropic anchoring regime}. For strong anchoring a sudden drop corresponds to the dynamic transition from the drifting, stretched finger pattern of Fig. \ref{CF1_tiltedFinger} to the bubble-like conformation show in Fig. \ref{CF1_driftBubble}. The second dynamic transition to the Grandjean-like texture has no specific signature in the viscosity curve. \rev{There is no shear thinning in CF1 for weak homeotropic anchoring. The results are compared with a cholesteric helix that is oriented along the gap or x-direction (CH-X).}}
\end{center}
\end{figure}

\begin{figure*}[htpb]
\begin{center}
\includegraphics[width=0.245\textwidth]{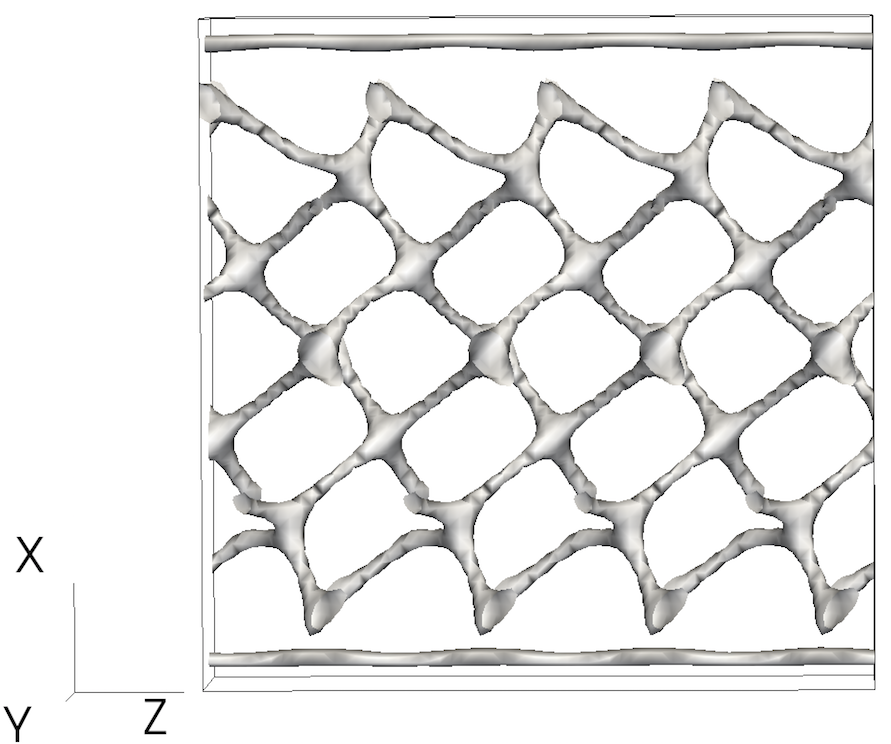}
\includegraphics[width=0.245\textwidth]{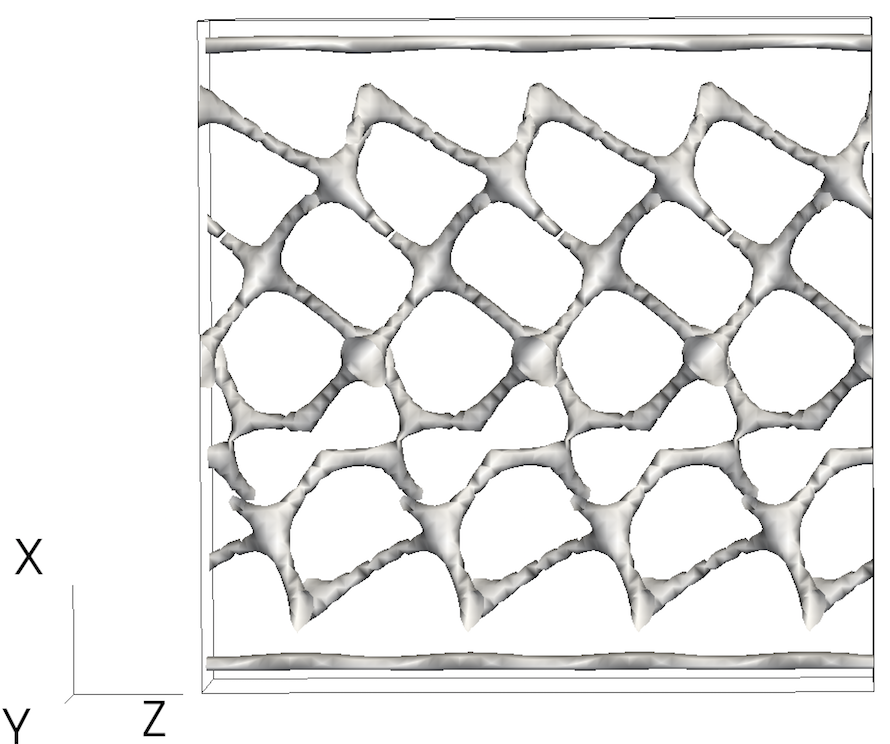}
\includegraphics[width=0.245\textwidth]{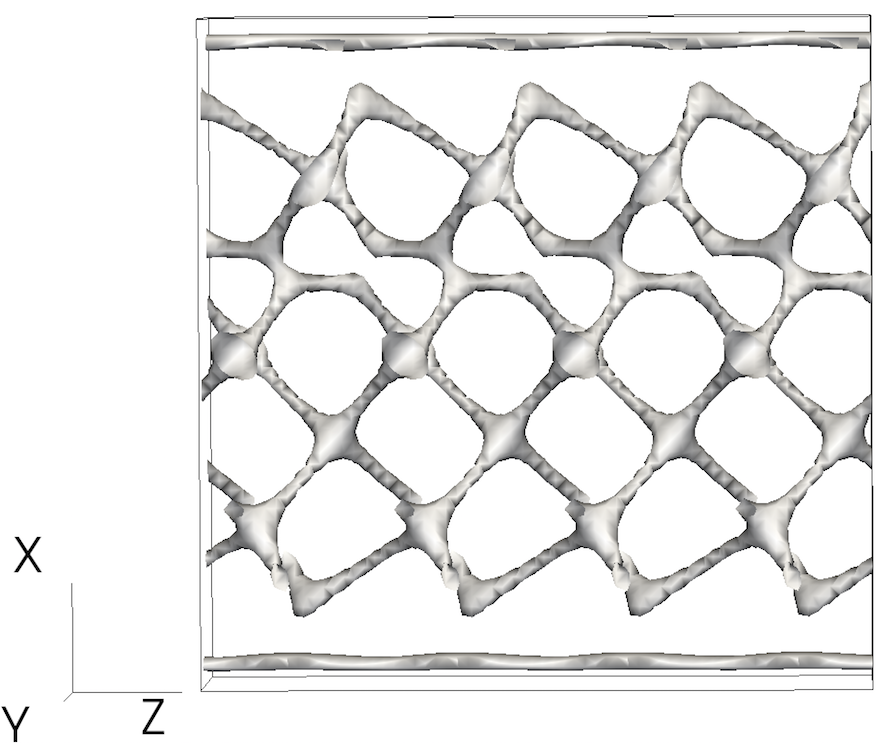}
\includegraphics[width=0.245\textwidth]{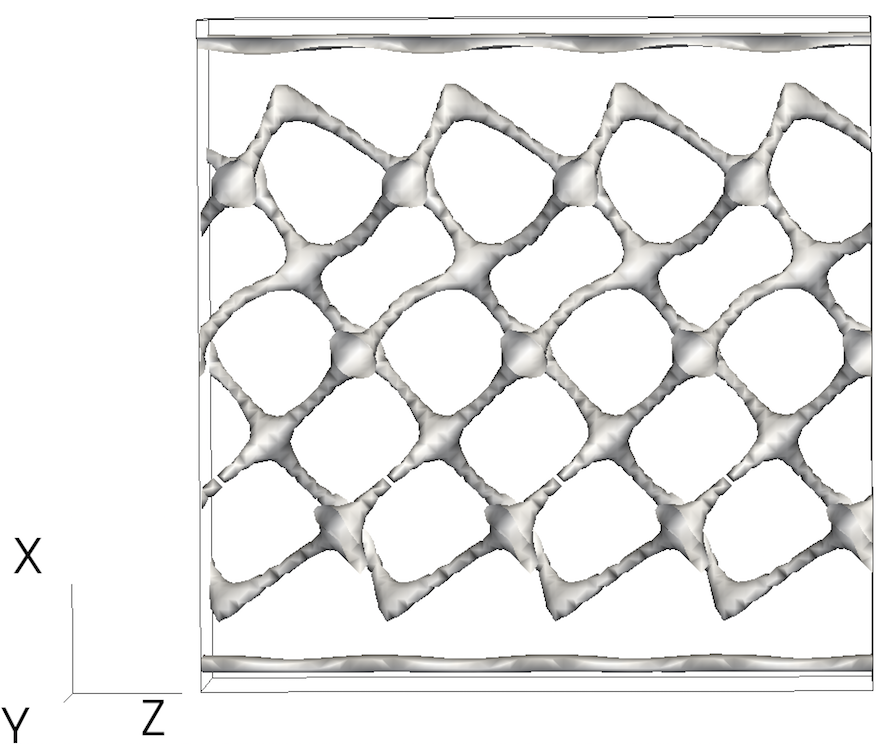}\\
\includegraphics[width=0.245\textwidth]{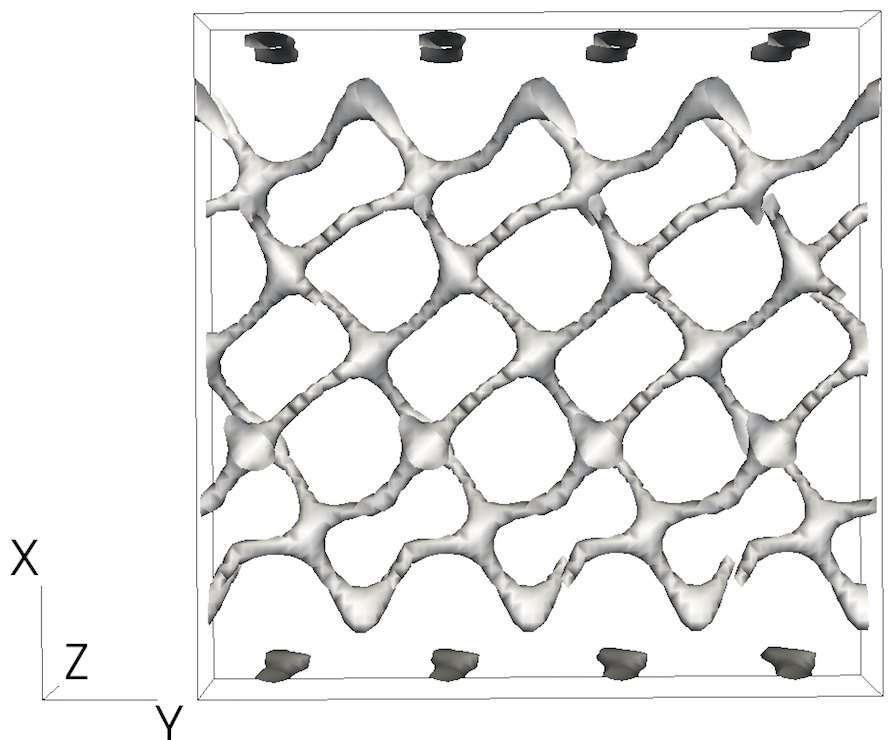}
\includegraphics[width=0.245\textwidth]{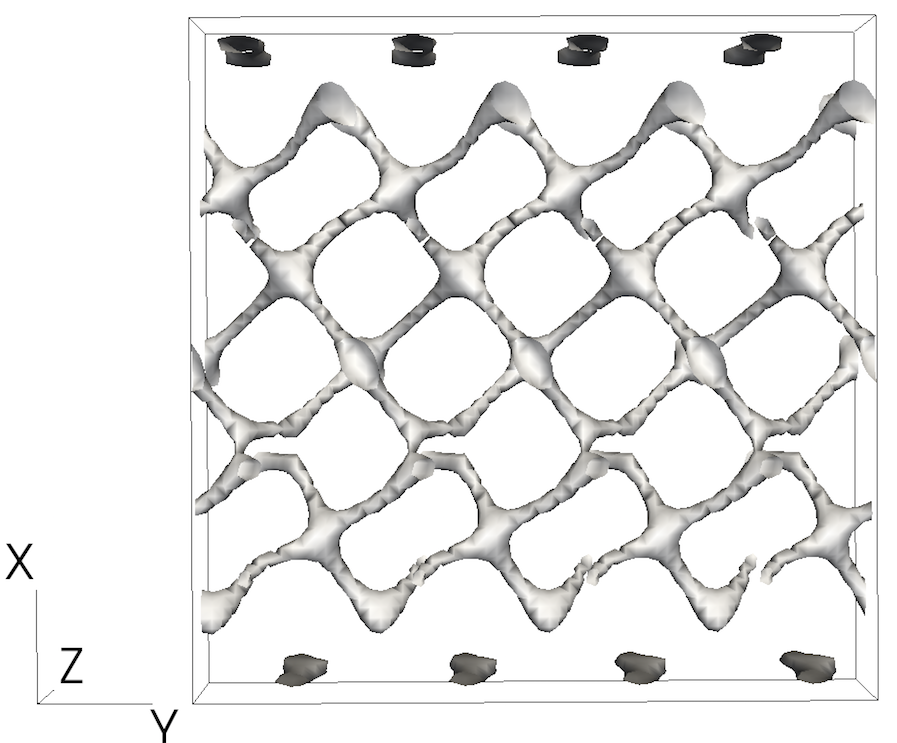}
\includegraphics[width=0.245\textwidth]{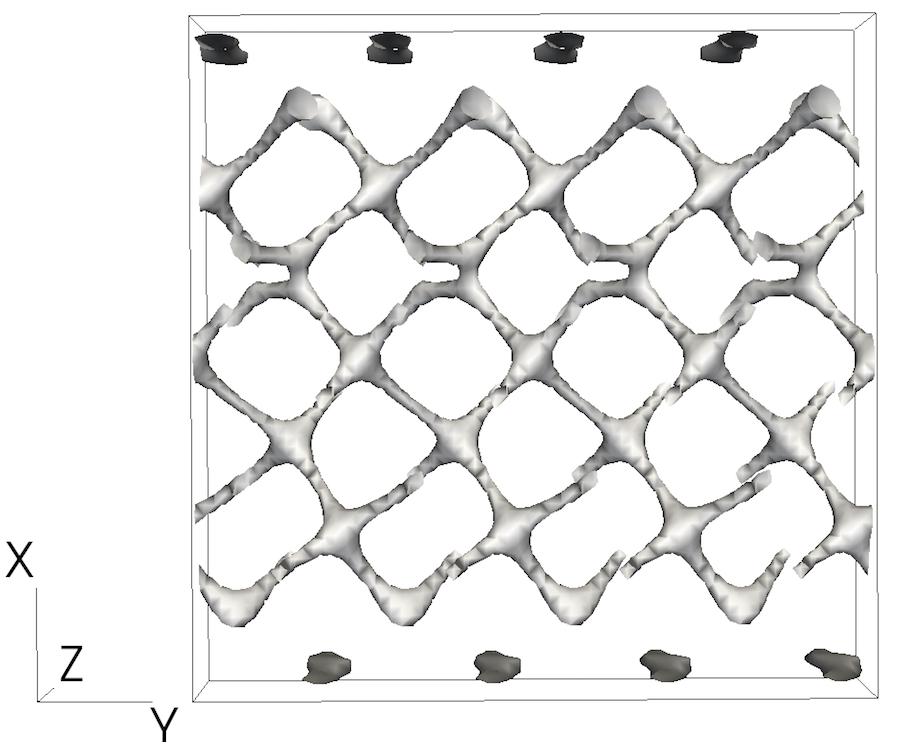}
\includegraphics[width=0.245\textwidth]{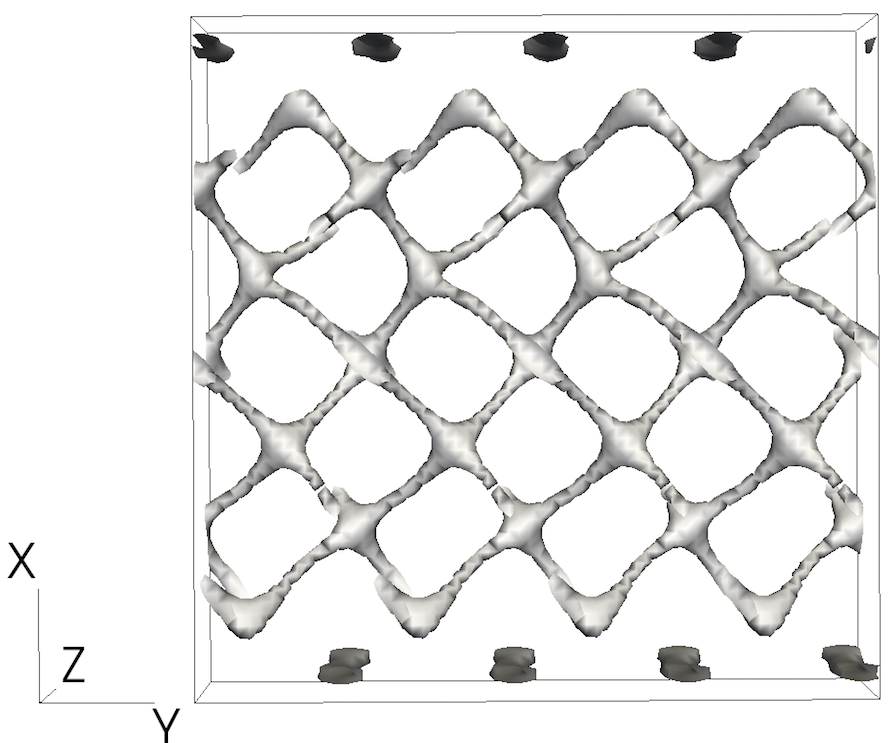}
\caption{\label{BP2_disc} Time sequence of the disclination network in flow. The images show a complete cycle of breakup and reformation of the disclination network at a forcing parameter $f=10^{-5}$ and time steps $t=161k, 170k, 178k$ and $184k$ (from left to right), respectively. The top row gives the view in y- or vorticity direction, whereas the bottom row depicts the network in z- or flow direction. The walls are at the top and bottom boundary along the x-direction.}
\end{center}
\end{figure*}

At larger forcing $f>5\times10^{-5}$ these bubble-like domains break up and the system approaches a standard cholesteric conformation in the steady state. This is shown in Fig. \ref{CF1_bands} for a pressure gradient of $f=10^{-4}$. In the centre of the channel the conformation is similar to a Grandjean texture and forms a chiral nematic region with the helical axis perpendicular to the flow direction. A similar state has been previously observed in simple shear ~\cite{Henrich:2013}. Closer to the walls the director field is affected by the homeotropic anchoring conditions and prefers alignment along x. This mode of flow has been observed up to pressure gradients of $f=5\times 10^{-4}$. While larger values cannot be feasibly studied with our algorithm, as they would violate the low Mach number requirement of the lattice Boltzmann method, we expect that the Grandjean-like texture will eventually unwind at higher flow rates into a flow-aligned nematic liquid crystal, as in~\cite{Henrich:2013}. 

The different dynamic regimes we identified above have a direct influence on the apparent viscosity of the flowing liquid crystal. Fig.\ref{CF1_viscosity} shows the apparent viscosity $\nu_{app}$  (computed according to Eqs.~\ref{volumetric_flow_rate} and \ref{volumetric_flow_rate_0}) over the pressure gradient $f$ \rev{in the weak and strong homeotropic anchoring limit. For strong anchoring (CF1, $W/K=10$)} the transition from the drifting finger pattern depicted in Fig. \ref{CF1_tiltedFinger} to the bubble-like conformation shown in Fig. \ref{CF1_driftBubble} results in a sudden drop of the apparent viscosity. This feature can be understood by closer inspection of the corresponding director profile. When the finger pattern detaches from the bottom wall, bubble-like domains form at the centre of the channel. These domains are free from anchoring effects as coupling with the walls would otherwise create a torque on the director field. Hence the liquid crystal can flow with less hindrance, \rev{leading to a state with reduced dissipation, shear stresses and apparent viscosity $\nu_{app}$.  The second dynamic transition, from the bubble-like domains to the Grandjean texture, does not entail a noticeable drop in the apparent viscosity as the dissipation in the bubble regime is already quite small. It has been previously shown ~\cite{Henrich:2013} that the dissipation in this Grandjean-like state is greatly reduced. Hence it comes as no surprise that this leads to viscosities that are very close to the background viscosity. 
It is interesting to see that in the strong anchoring regime a simple cholesteric helix oriented along the x-direction (CH-X, $W/K=10$) shows a gradual decrease of the apparent viscosity, which is lower than that of CF1 at small pressure gradients, but slightly higher for large pressure gradients. In the weak anchoring case both phases CF1 and CH-X show no shear thinning at all as the coupling with the wall is already greatly reduced. Hence, we observe small viscosities throughout.}

\subsection{Blue Phases }

\begin{figure*}[htpb]
\begin{center}
\includegraphics[width=0.45\textwidth]{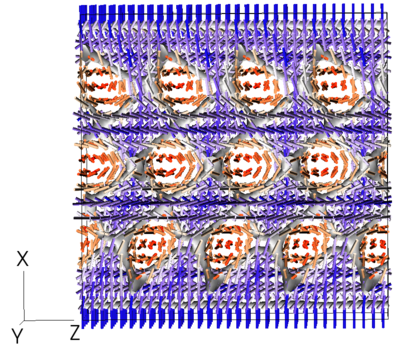}
\includegraphics[width=0.467\textwidth]{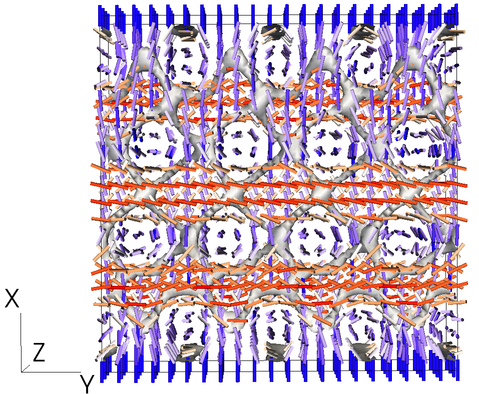}\\
\includegraphics[width=0.45\textwidth]{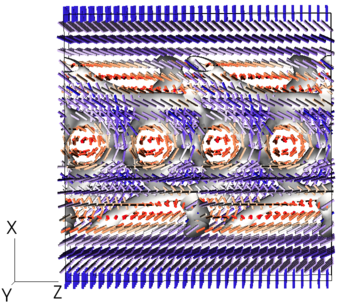}
\includegraphics[width=0.461\textwidth]{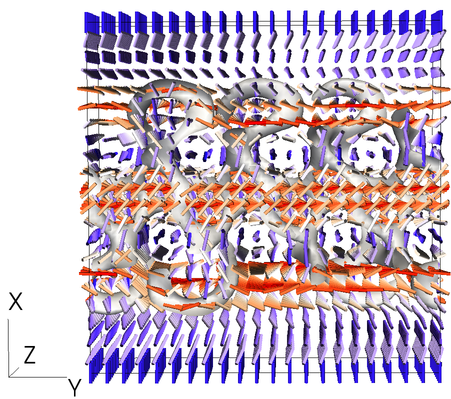}\\
\includegraphics[width=0.45\textwidth]{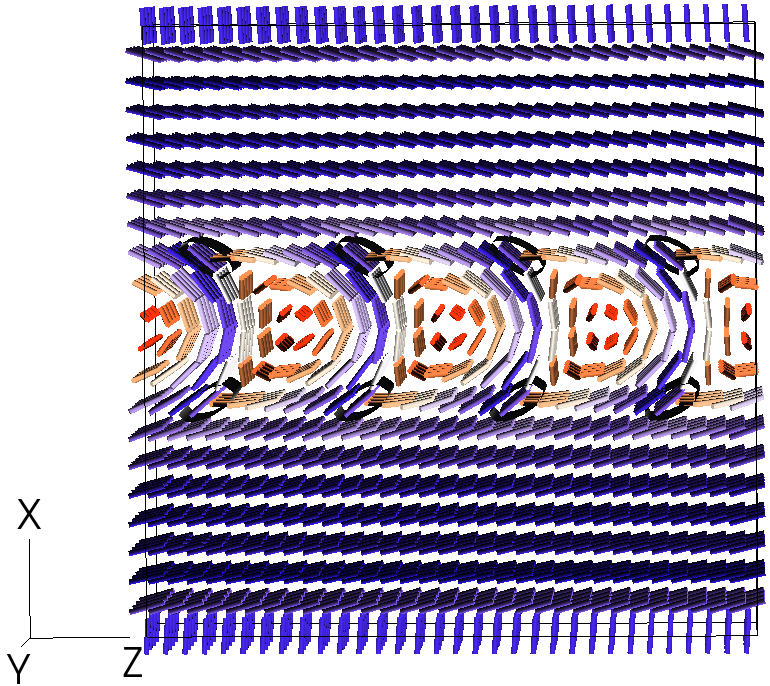}
\includegraphics[width=0.461\textwidth]{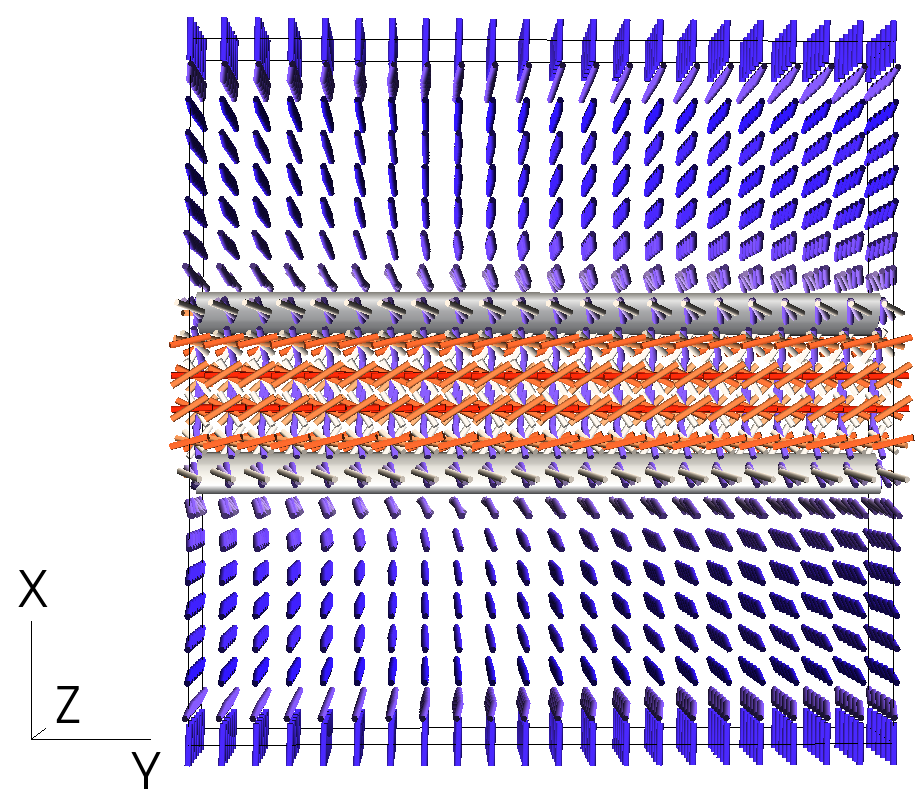}
\caption{\label{BP2_norm_w4e-1} The effect of an increasing pressure gradient on the double twist cylinder (DTC) structure for strong homeotropic anchoring conditions. In the left column the flow is along the horizontal axis from left to right, whereas in the right column the flow is into the plane. At $f=1\times10^{-5}$ (top row) DTCs resist the flow. At $f=1\times10^{-4}$ (middle row) the DTCs closest to the wall disappear, whereas the remaining cylinders continue to flow. At even larger pressure gradients of $f=5\times10^{-4}$ (bottom row) all but the central row of DTCs has been destroyed. At the top and bottom walls homeotropic strong anchoring conditions have been imposed with $W/K=10$.}
\end{center}
\end{figure*}

\begin{figure*}[htpb]
\begin{center}
\includegraphics[width=0.45\textwidth]{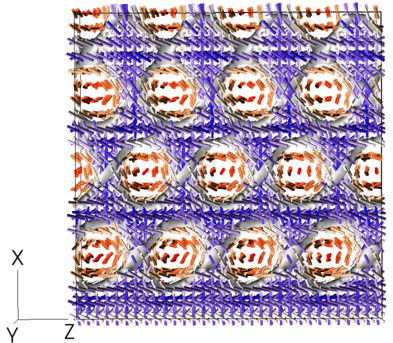}
\includegraphics[width=0.467\textwidth]{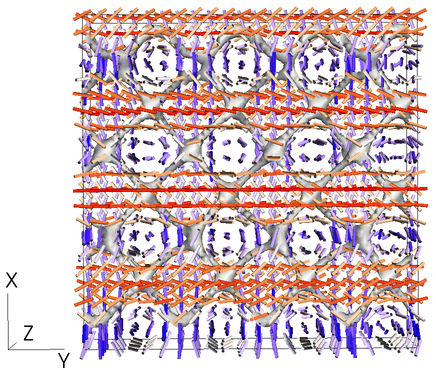}
\caption{\label{BP2_norm_w4e-3} The case of weak homeotropic anchoring conditions ($W/K=0.1$) and a representative pressure gradient of $f=1\times10^{-5}$. In the left picture the flow is along the horizontal axis from left to right whereas in the picture on the right the flow is into the plane. The director field looks almost identical to the shown in Fig.\ref{BP2_norm_w4e-1}, but due to the weaker anchoring strength the director field is not forced into to align perpendicular to the walls.}
\end{center}
\end{figure*}

\begin{figure}[htpb]
\begin{center}
\includegraphics[width=0.495\textwidth]{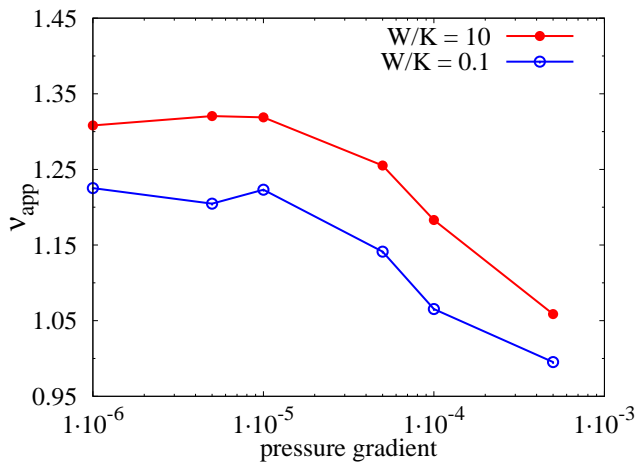}
\caption{\label{BP2_norm_viscosity} Apparent viscosity $\nu_{app}$ against forcing parameter $f$ for BPII with weak and strong homeotropic anchoring conditions at the walls.}
\end{center}
\end{figure}

Contrary to Cholesteric Fingers, Blue Phases (BPs) are genuinely three-dimensional structures and cannot be modelled in a quasi-2D geometry. As BPs are known to show a very intricate flow behaviour that depends sensitively on the local topology of the disclination network, we limited our studies to Blue Phase II (BPII) only, which has a somewhat simpler response to flow than Blue Phase I (BPI)~\cite{Henrich:2012b,Henrich:2013}. In the following pictures the system dimensions are $L_x \times L_y \times L_z = 64^3$. For clarity, we sometimes show only a subsection of the entire system or prune the director field. 
We modelled two different anchoring conditions and strengths, namely weak or strong homeotropic or planar degenerate anchoring with a fixed anchoring strength at the walls. In order to generate a thermodynamically equilibrated BP structures it was necessary to divide the equilibration phase into two separate stages. First, we undertook a short run with fully periodic boundary conditions during which the redshift was dynamically adapted. This was followed by the second stage where we introduced solid walls and the wall anchoring. After this second equilibration phase we extracted the value of the redshift from the simulations and used it as fixed input parameter for the runs with pressure gradients. 
In terms of applied pressure gradients a similar range of values was used with a slightly lower resolution than for the CF1 due to increased computational costs.

\subsubsection{Homeotropic Anchoring }\label{bp-norm}

Similar to what has been observed for the CF1, the BPII network does not move visibly at pressure gradients below $f\lesssim 5\times 10^{-7}$. However, a finite mass flow was measured, which is again a consequence of permeative flow at these very low pressure gradients. At larger pressure gradients the disclination network cannot withstand the strain and begins to move downstream by breaking up and reforming periodically. \rev{This cyclic behaviour gives rise to the previously reported  oscillatory stress patterns ~\cite{Dupuis:2005, Henrich:2013}}. Snapshots of this movement are shown in in Fig. \ref{BP2_disc}, for a representative pressure gradient of $f=10^{-5}$. During the 23k simulation timesteps that the sequence comprises the structure at the centre travels about 4 unit cells downstream and roughly once through the entire simulation box. 

Interestingly, the disclination network does not only move downstream along the z-direction, but travels also, albeit more slowly, along the vorticity or y-direction. This peculiar behaviour has been previously observed in simple shear flow of bulk BPs ~\cite{Henrich:2013} and also partly in simple shear flow of BPs that are confined between two walls with pinned boundary conditions ~\cite{Henrich:2012b}. It could be directly linked to (a) the orientation of the velocity gradient and (b) the helicity of the conformation of the underlying cholesteric liquid crystal.

\begin{figure}[htpb]
\begin{center}
\includegraphics[width=0.49\textwidth]{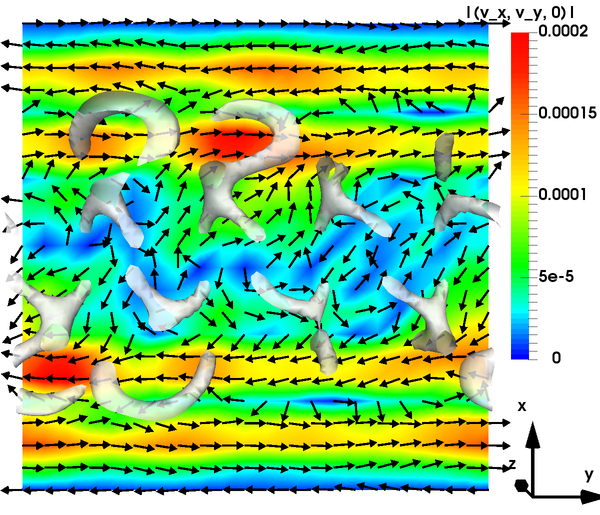}\\
\includegraphics[width=0.49\textwidth]{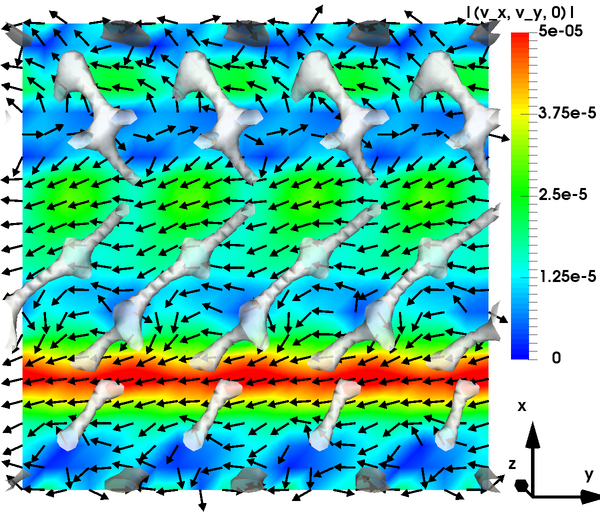}
\caption{\label{BP2_norm_secondary_flow} \rev{Slices through the secondary flow velocity pattern $(v_x, v_y, 0)$ in xy-plane at $L_z= p/4$ for strong homeotropic anchoring. The pictures display snapshots at time step $t=250k$ for a pressure gradient of $f=10^{-4}$ (top) and $f=10^{-5}$ (bottom), respectively. The colour code gives the magnitude of the secondary flow velocity, i.e. the flow velocity with omitted primary flow component $v_z$ (see also Tab. \ref{table1}), whereas the arrows show the direction of the secondary flow field in xy-plane. The grey isosurfaces show the disclination network immediately before the intersection plane.}}
\end{center}
\end{figure}

Differently than in simple shear flow, in a Poiseuille flow the velocity gradient is not uniform, but spatially inhomogeneous. It peaks at or very close to the walls, and decreases towards the centre of the gap. It also has different signs on either side of the central plane. Consequently, the network moves along the negative y-direction in the top half of the channel, and along the positive y-direction in the bottom half. This is clearly visible in the sequence of snapshots in the bottom row of Fig. \ref{BP2_disc}. Remarkably, this phenomenon leads to \rev{a complex secondary flow pattern in the velocity components perpendicular to the primary flow direction, i.e. $v_x$ and $v_y$ in our case. This secondary flow is typically up to two orders of magnitude smaller than the primary flow, and changes direction during the cyclic breakup and reformation of the disclination network (see also Fig. \ref{BP2_norm_secondary_flow} and discussion below.)} 

\begin{figure*}[htpb]
\begin{center}
\includegraphics[width=0.495\textwidth]{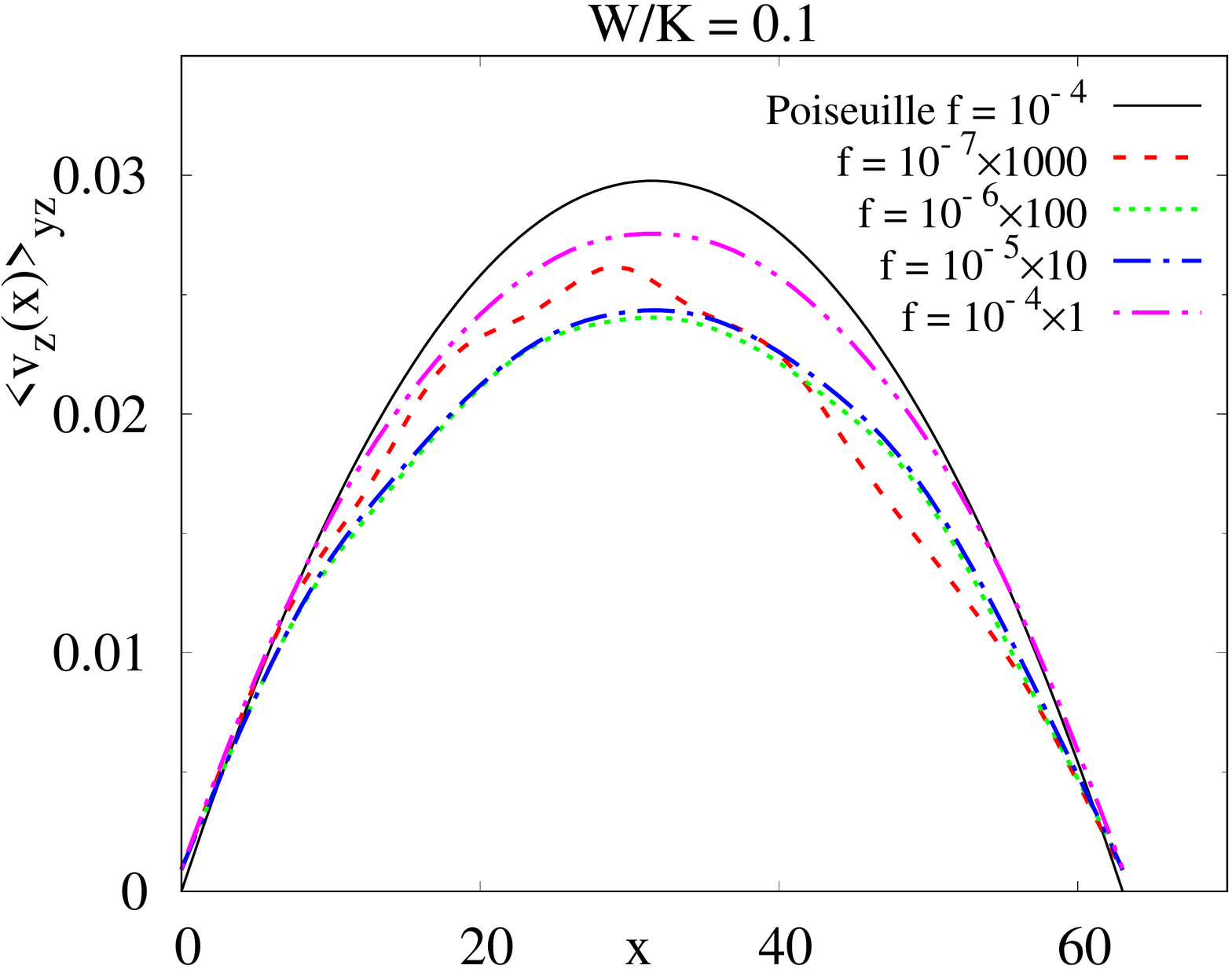}
\includegraphics[width=0.495\textwidth]{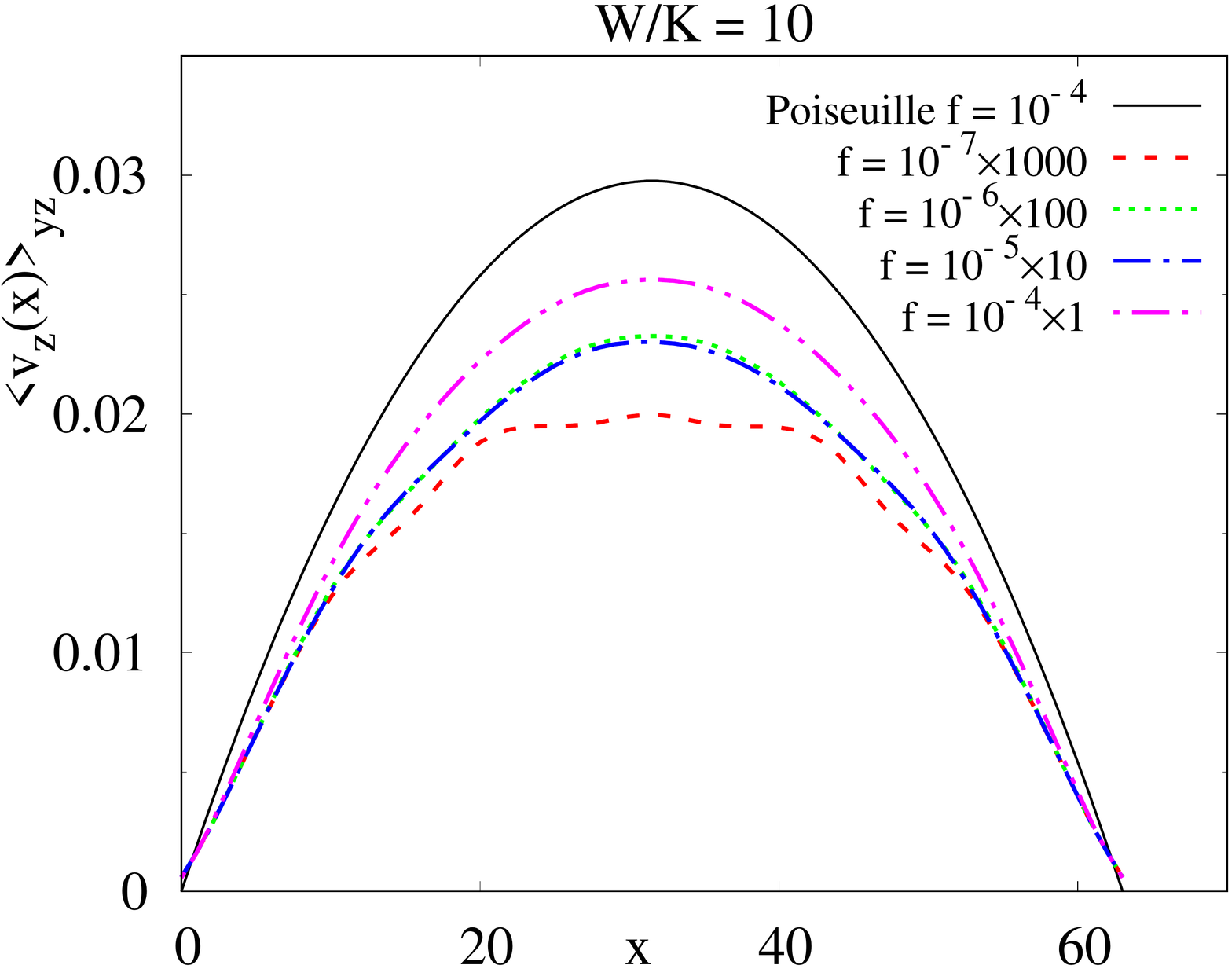}\\
\includegraphics[width=0.495\textwidth]{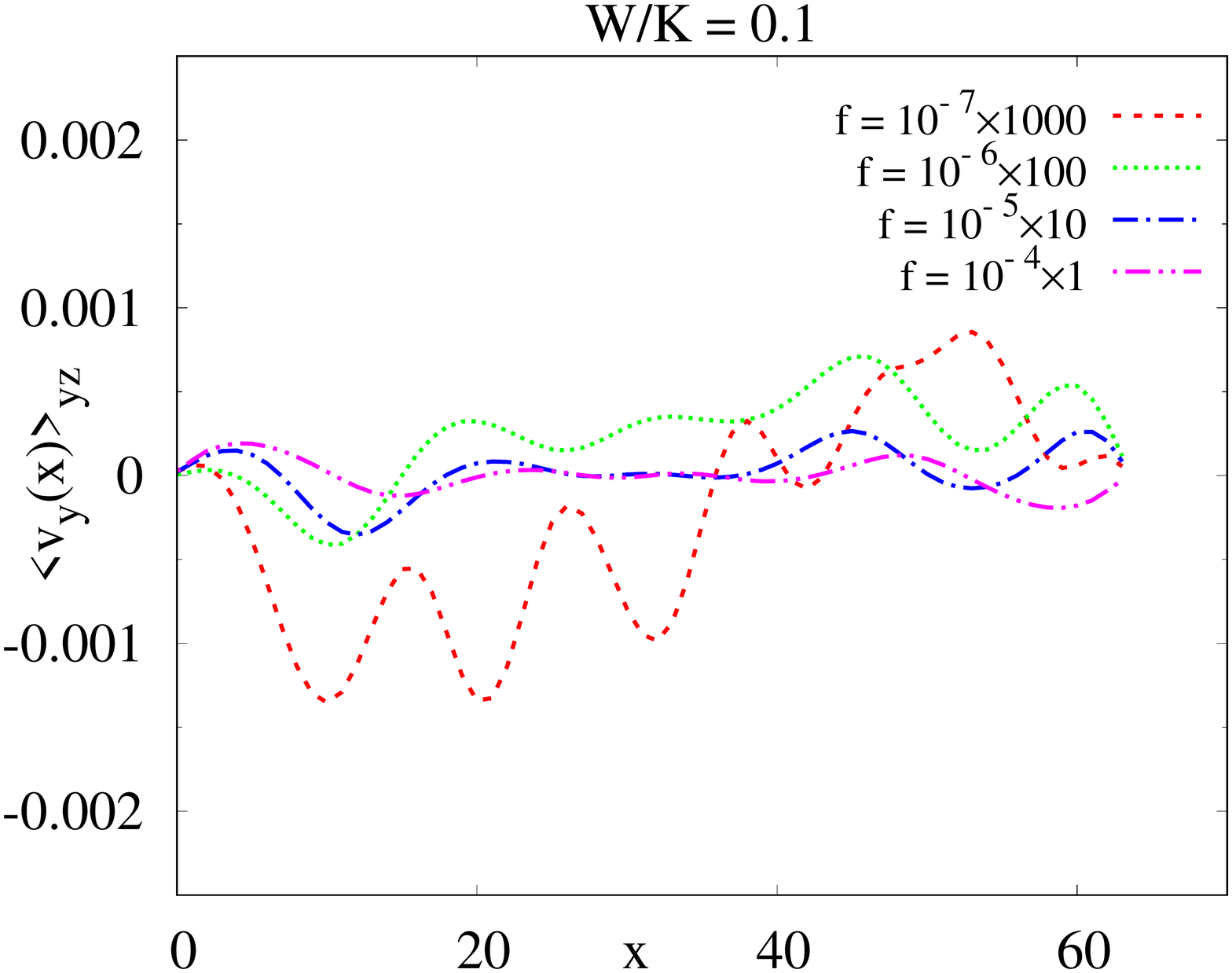}
\includegraphics[width=0.495\textwidth]{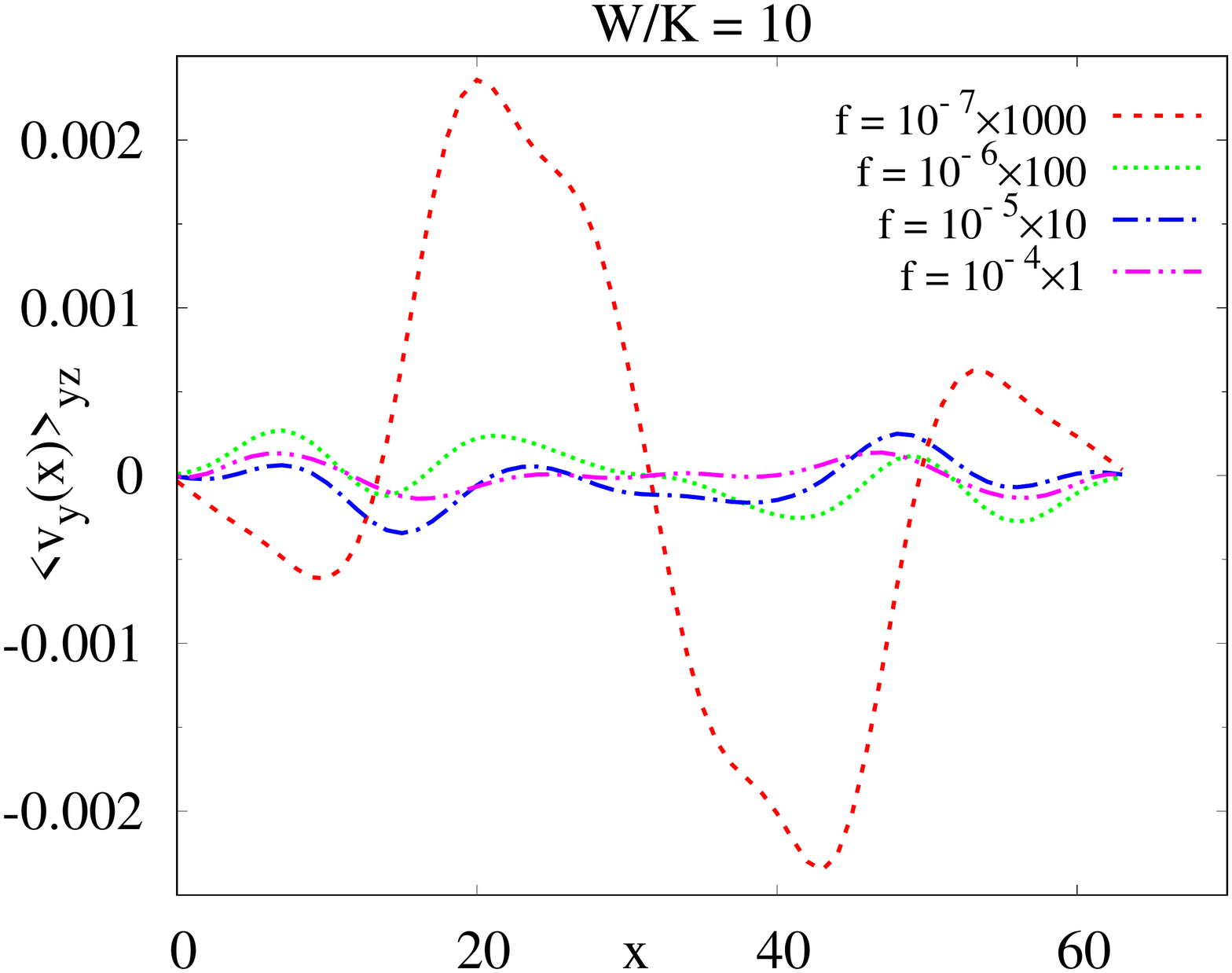}\\
\caption{\label{BP2_norm_velocity} Flow velocity profiles at different pressure gradients $f$ for weak (left column) and strong (right column) homeotropic anchoring. The top row shows snapshots of the z-component of the flow velocity $\langle v_z(x)\rangle_{yz}$ across the gap in x-direction (averaged along the y and z direction). The bottom row gives snapshots of the y-component of the flow velocity $\langle v_y(x)\rangle_{yz}$ (averaged along the y and z direction).  Note that the data has been scaled to make it comparable to the data obtained for the largest applied pressure gradient.} 
\end{center}
\end{figure*}

A closer look at the director field for different pressure gradients and strong homeotropic anchoring conditions at the walls ($W/K=10$) is given in Fig.\ref{BP2_norm_w4e-1}.  \rev{The pictures show the director field with the colour code displaying the x-component (perpendicular to the wall) of the director field ranging. As previously, the colour ranges from blue (director perpendicular to the wall) over light blue, white and orange to red (director parallel to the wall). The light grey isosurfaces depict regions where the scalar order parameter drops below a value of $q_s\simeq0.12$. Clearly visible are regions of horizontally oriented DTCs, indicated by the red and orange colour of the director field.} It is obvious that, as the pressure gradient increases, the DTC structure of the BP persists only at the centre of the channel. Interestingly, they actually remain virtually intact throughout the pattern of breakup and reformation of the disclination network shown in Fig. \ref{BP2_disc}. This is because DTCs are the preferred local thermodynamic configuration for the chosen parameter set. Closer to the walls, the velocity gradients are large, and these dissolve the near-wall DTCs gradually: the LC adopts a flow-aligned Grandjean-like texture in these regions. (Only in the direct vicinity of the walls the anchoring is strong enough to enforce normal orientation of the director field.) 
At even larger pressure gradients above $f=5\times10^{-4}$ the DTCs disappear altogether and a Grandjean-like texture similar to the one shown in Fig. \ref{CF1_bands} prevails throughout the entire system with the exception of the regions very close to the walls. 

\begin{table*}[htpb]
\begin{center}
\begin{tabular}{ | l | l | l | l | l |}
\hline
$f$ & Min/max secondary velocity $v_x$ & Min/max secondary velocity $v_y$ & Min/max primary velocity $v_z$ & $Er$\\ \hline 
$10^{-4}$ & \rev{$-1.01\times10^{-4}$ / $ 9.35\times10^{-5}$} & $-2.16\times10^{-4}$ / $ 2.18\times10^{-4}$ &  $5.66\times10^{-4}$ / $ 2.57\times10^{-2}$ & 21.905\\  %
$10^{-5}$ & \rev{$-2.48\times10^{-5}$ / $ 2.70\times10^{-5}$} & $-5.63\times10^{-5}$ / $ 4.49\times10^{-5}$ &  $5.55\times10^{-5}$ / $ 2.31\times10^{-3}$ & 1.972\\%
$10^{-6}$ & \rev{$-8.20\times10^{-6}$ / $ 8.13\times10^{-6}$} & $-1.93\times10^{-5}$ / $ 1.94\times10^{-5}$ &  $2.82\times10^{-6}$ / $ 2.34\times10^{-4}$ & 0.199\\ %
$10^{-7}$ & \rev{$-8.04\times10^{-6}$ / $ 8.00\times10^{-6}$} & $-1.91\times10^{-5}$ / $ 1.90\times10^{-5}$ &  $-2.53\times10^{-6}$ / $ 2.81\times10^{-5}$ &0.024\\ %
\hline
\end{tabular}
\end{center}
\caption{\label{table1} \rev{The secondary velocity components $v_x$ and $v_y$ and the primary flow component $v_z$ for strong homeotropic anchoring conditions and different pressure gradients $f$. The data gives the minimum and maximum value of the components and the Ericksen number $Er$ during a snapshot at time step $t=250k$.}}
\end{table*}   	

The situation is similar for weak homeotropic anchoring with $W/K=0.1$, shown in Fig.~\ref{BP2_norm_w4e-3} for a representative pressure gradient of $f=1\times10^{-5}$. A distinctive array of DTCs is visible, but minor differences occur very close to the walls. Due to the weaker anchoring strength the director field can be rotated away from the anchoring-induced orientation by thermodynamic or flow-induced torques. As a result, and in contrast with the case of strong homeotropic anchoring depicted in Fig. \ref{BP2_norm_w4e-1}, a fourth row of partial DTCs forms along the y-direction at the very top of the system. This fact has a measurable influence on the apparent viscosity of the flowing BP.   

Fig.~\ref{BP2_norm_viscosity} shows the apparent viscosity following Eqs.\ref{volumetric_flow_rate} and \ref{volumetric_flow_rate_0} for weak and strong anchoring with $W/K=0.1$ and $W/K=10$, respectively. The system with stronger normal anchoring clearly creates a larger resistance against the pressure gradient leading to a higher apparent viscosity. However, for both anchoring strengths the curves follow roughly the same trend with a plateau-like region at intermediate pressure gradients followed by a decreasing apparent viscosity to reach $\nu_0$ at large pressure gradients, corresponding to shear thinning. Comparing the whole curve for BPII and CF1, we find two main differences: first, BPII has a smaller apparent viscosity at low forcing; second, for strong homeotropic anchoring the decline in viscosity is a lot smoother for BPII, which is directly related to the rather gradual breakup of the DTCs with increasing pressure gradient, which can be seen in Fig. \ref{BP2_norm_w4e-1}. 

The anchoring strength and the disclination network impacts on the flow velocities as well. Fig. \ref{BP2_norm_velocity} shows averaged velocity profiles of the BPs for weak and strong homeotropic anchoring. The top row shows y- and z-averaged snapshots of the z-component of the flow velocity $\langle v_z(x)\rangle_{yz}$ at particular times during the flow cycles. 
The data has been scaled by the magnitude of each pressure gradient to allow direct comparison of the data for all pressure gradients in one picture. Clearly visible is that for larger pressure gradients the dominating velocity component in z-direction approaches the parabolic Poiseuille profile, here representatively plotted for a pressure gradient of $f=10^{-4}$. For intermediate pressure gradients the liquid crystal instead flows more in a plug-like manner. At the lowest pressure gradient $f=10^{-7}$ where we observe permeative flow, we detect small fluctuations, a residual effect of the stationary DTC structure. In the case of weak normal anchoring there is also an asymmetry, a consequence of the fact that the DTCs are not completely symmetrically aligned in the channel, as e.g. visible in Fig. \ref{BP2_norm_w4e-3}.    

The bottom row of Fig. \ref{BP2_norm_velocity} shows y- and z-averaged snapshots of the y-component of the flow velocity $\langle v_y(x)\rangle_{yz}$ across the gap. These are generally at least one, \rev{but up to two} orders of magnitude smaller than the primary velocity component $v_z$ along the flow direction.
This secondary flow is induced by the underlying cholesteric ordering and occurs along the directions perpendicular to the primary flow direction. The spatial dependency of these averages is less clear, but note that these profiles represent only a snapshot during a specific time during one cycle of breakup and reformation of the disclination network. There is an overall trend of these secondary velocity components to become smaller at larger pressure gradients as the DTC structure and disclination network gradually breaks up and the LC adopts the Grandjean-like texture in the regions with large velocity gradients (see Fig.\ref{BP2_norm_w4e-1} bottom row). 

\rev{In Fig. \ref{BP2_norm_secondary_flow} we show cuts through the secondary flow velocity pattern at $L_z=p/4$, i.e. a quarter of a pitch length away from the system boundary. The cuts were obtained by projecting the total flow velocity $\vec{v}=(v_x, v_y, v_z)$ onto xy-planes perpendicular to the primary flow direction (and hereby projecting out the dominating $v_z$-component). The resulting projected velocity vectors are then visualised with their direction in xy-plane (arrows) and their magnitude (colour code). Although the magnitude is in the percent range of the primary flow velocity, it is interesting to see that there are very clearly defined regions where the liquid crystal flows to a small degree also perpendicularly to the primary flow. For the larger pressure gradient $f=10^{-4}$ we see actually counterflow in both the top and the bottom half of the channel. It is worth mentioning that these patterns change constantly as the disclination network breaks up and reconnects as it flows downstream.}

Table \ref{table1} shows an overview of the extreme values of the primary and secondary flow components at a specific time during a cycle of breakup and reformation of the disclination network and how these vary with the applied pressure gradient. Quite strikingly, it can be seen that the minimum values of the primary flow changes by almost two orders of magnitude, and change even sign for $f=10^{-7}$. As for the secondary flow velocity components $v_x$ and $v_y$, their magnitude reaches that of the primary flow velocity component $v_z$ for the lowest applied pressure gradients.

\subsubsection{Planar Anchoring }\label{bp-plan}

\begin{figure*}[htpb]
\begin{center}
\includegraphics[width=0.45\textwidth]{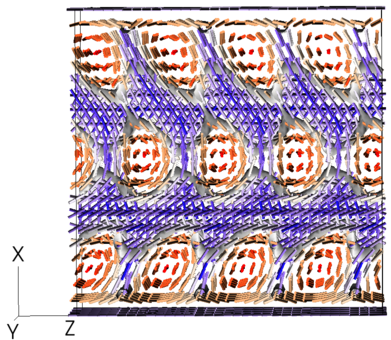}
\includegraphics[width=0.47812\textwidth]{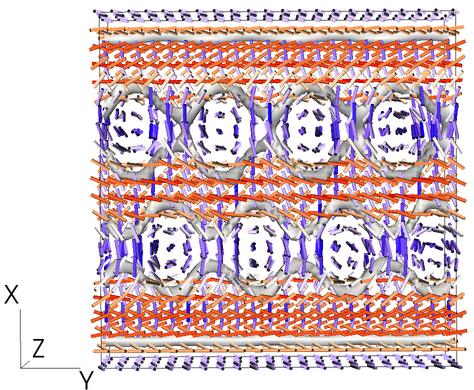}\\
\includegraphics[width=0.45\textwidth]{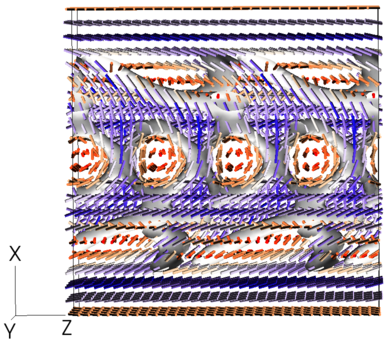}
\includegraphics[width=0.47812\textwidth]{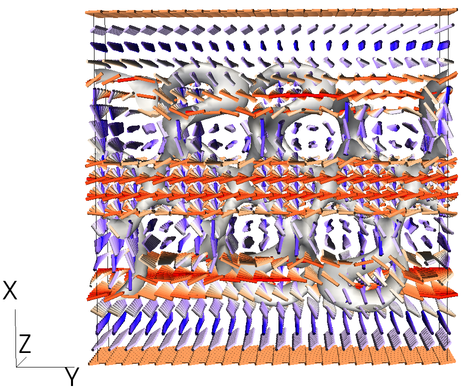}\\
\includegraphics[width=0.45\textwidth]{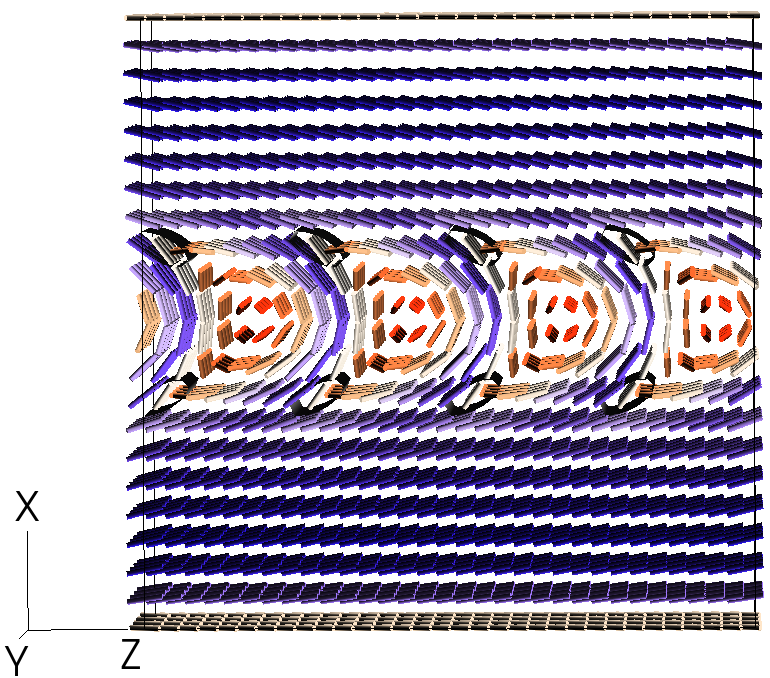}
\includegraphics[width=0.4725\textwidth]{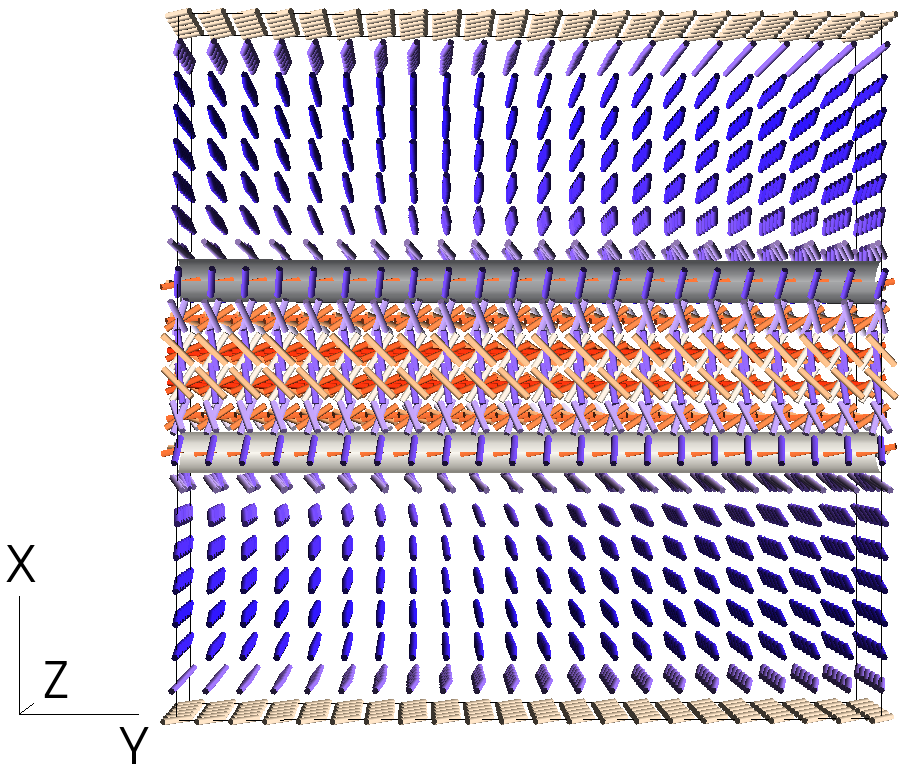}
\caption{\label{BP2_plan_w4e-1} Effect of increasing pressure gradient on the double twist cylinder (DTC) structure, for strong planar degenerate anchoring conditions. In the left column the flow is along the horizontal axis from left to right, whereas in the pictures on the right the flow is into the plane. At $f=1\times10^{-5}$ (top row) DTCs resist the flow. The array of DTCs appears slightly stretched with the top and bottom row of DTCs now situated closer to the walls than for homeotropic anchoring. At $f=1\times10^{-4}$ (middle row) the DTCs start to disappear beginning from the walls. The remaining cylinders continue to flow. At even larger pressure gradients of $f=5\times10^{-4}$ (bottom row) all but the central row of DTCs has been destroyed. At the top and bottom walls strong planar degenerate anchoring conditions have been imposed with $W_1/K=W_2/K=10$.}
\end{center}
\end{figure*}

We now turn to the case of planar degenerate anchoring conditions, for which the director field prefers to be aligned parallel to the wall surface. It can, however, freely rotate in plane without any energetic constraints or costs. On the level of the Landau-deGennes free energy functional this is modelled by introducing a fourth-order term ~\cite{Fournier:2005} (see Eq. \ref{planar_ac}).
Fig. \ref{BP2_plan_w4e-1} shows how the DTC structure is modified by the applied pressure gradient for anchoring strength $W_1/K=W_2/K=10$. This is the analogue situation of the one depicted in Fig. \ref{BP2_norm_w4e-1} for homeotropic anchoring. For simplicity we always set the two strengths characterising planar degenerate anchoring to be the same, i.e. $W_1=W_2$. 

\begin{figure}[htpb]
\begin{center}
\includegraphics[width=0.495\textwidth]{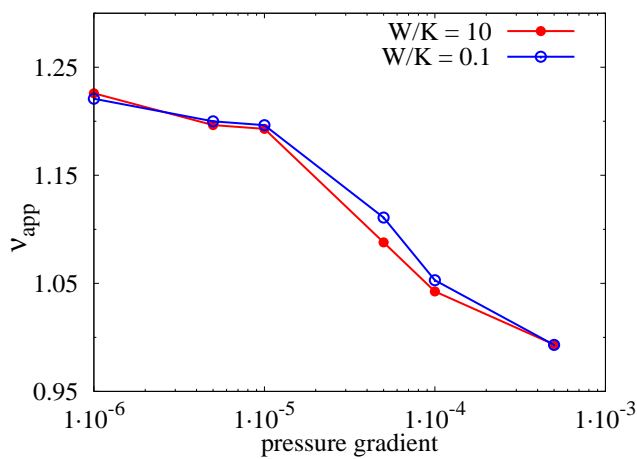}
\caption{\label{BP2_plan_viscosity} Apparent viscosity $\nu_{app}$ against pressure gradient $f$ for BPII with weak and strong planar degenerate anchoring conditions at the walls.}
\end{center}
\end{figure}

At the lower end of applied pressure gradients, as shown in the top row of Fig. \ref{BP2_plan_w4e-1} for $f=10^{-5}$, we find that planar degenerate anchoring conditions have a twofold effect on the conformation of the confined BPII.
First, the anchoring dictates a change in the preferred orientation of the liquid crystal in the immediate vicinity of the walls. Second, planar degenerate anchoring is more compatible with the director conformation inside DTCs; this leads to an effective attraction of DTCs to the wall, which results in a stretching of the BPI disclination network across the channel, with the DTC further apart with respect to the case of homeotropic anchoring case in Fig. \ref{BP2_norm_w4e-1}. The x-position of the y-oriented DTCs is a clear manifestation of this feature (see Fig.\ref{BP2_plan_w4e-1}, top left image).

This conformational difference between the two anchoring conditions becomes less pronounced as the forcing $f$ increases, as also for planar degenerate anchoring the DTCs begin to dissolve, starting from the walls and continuing towards the centre of the channel. 
Nevertheless, the behaviour in the proximity of the walls remains different across the whole range of pressure gradients. This is because planar anchoring conditions are fully compatible with flow alignment, whereas homeotropic anchoring prevents the director field from aligning with the flow close to the walls. While this may appear as a minor detail as it affects only a small layer of liquid crystal close to the wall, it has a measurable effect on the apparent viscosity of the confined BPII.

\begin{figure*}[htpb]
\begin{center}
\includegraphics[width=0.495\textwidth]{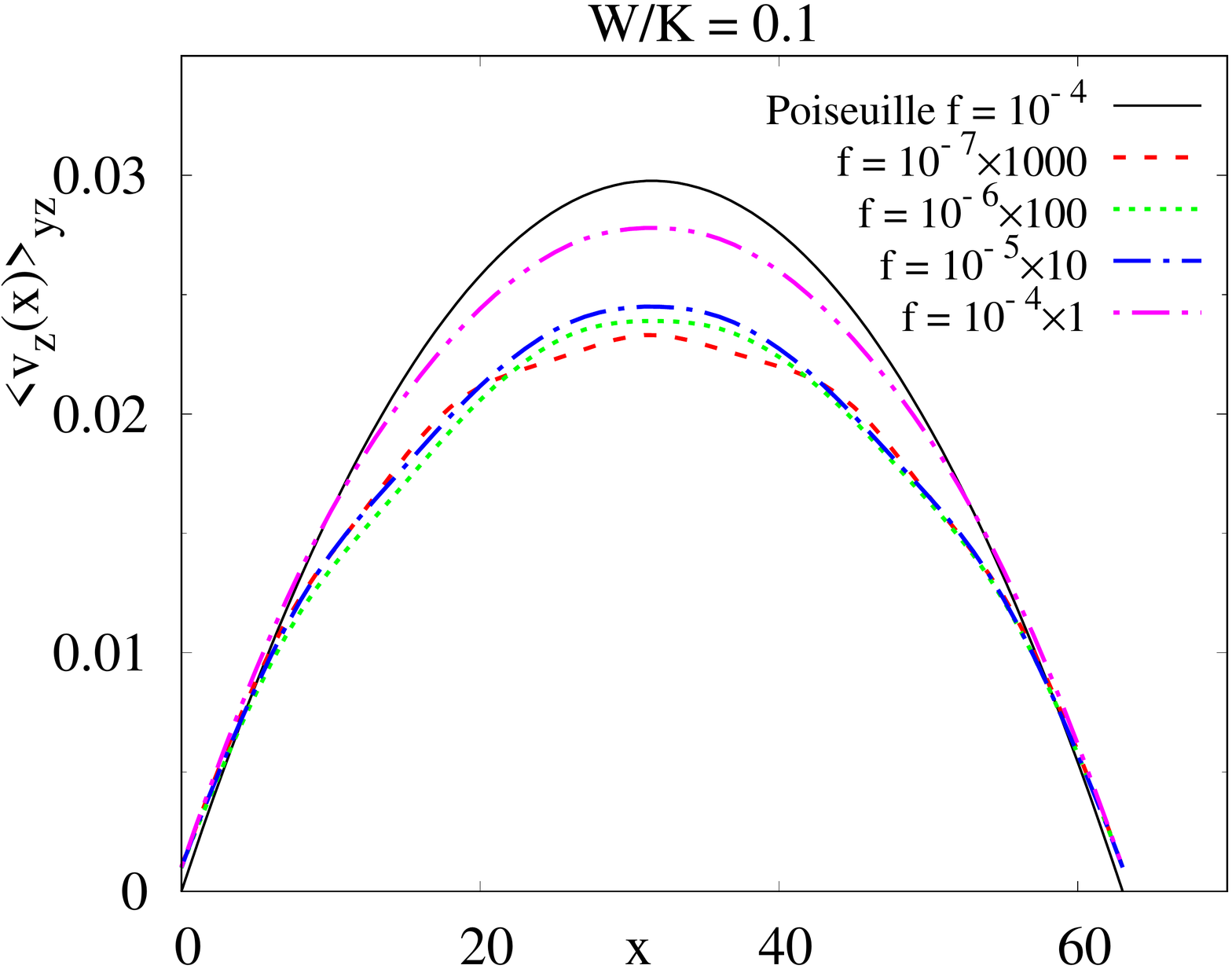}
\includegraphics[width=0.495\textwidth]{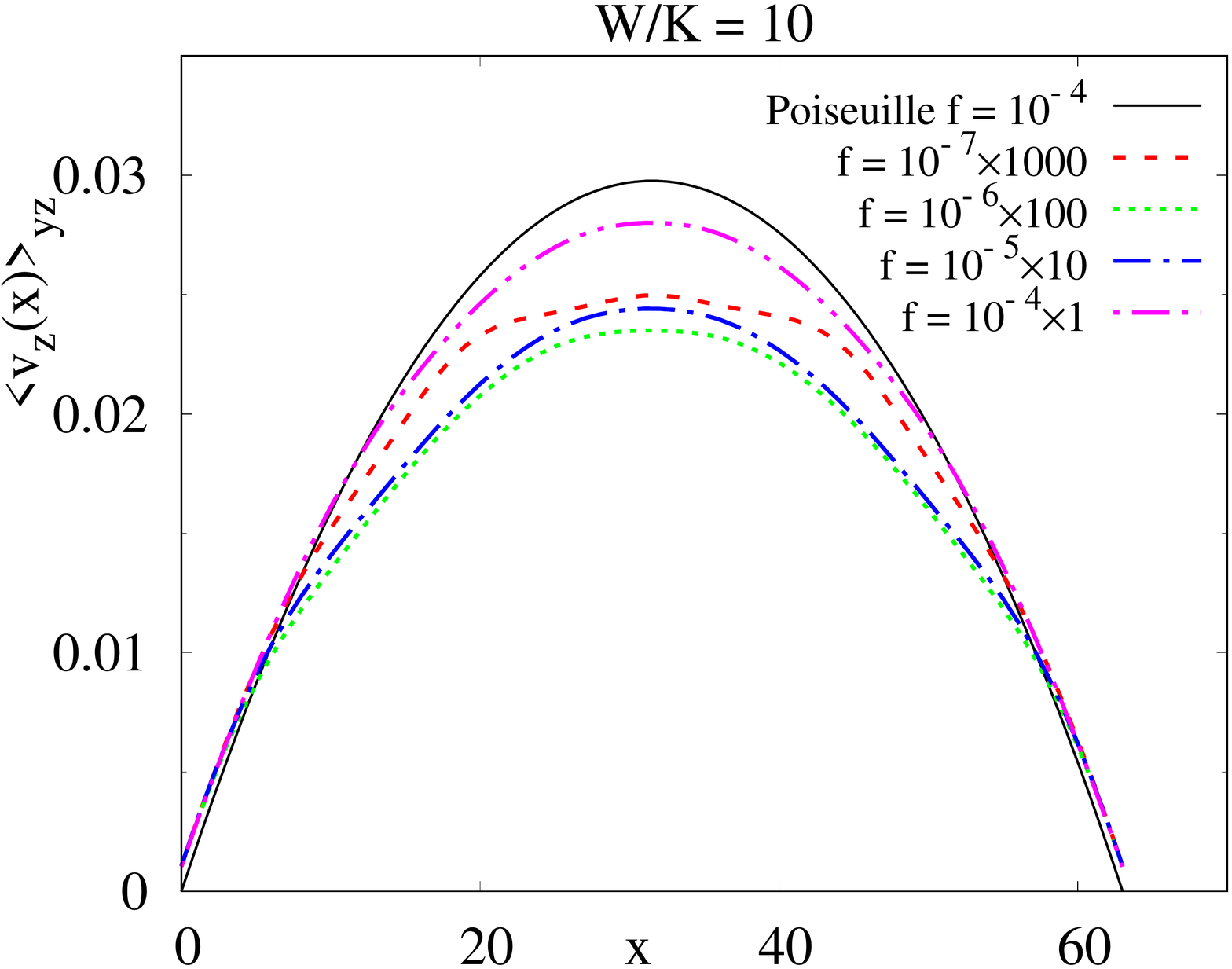}\\
\includegraphics[width=0.495\textwidth]{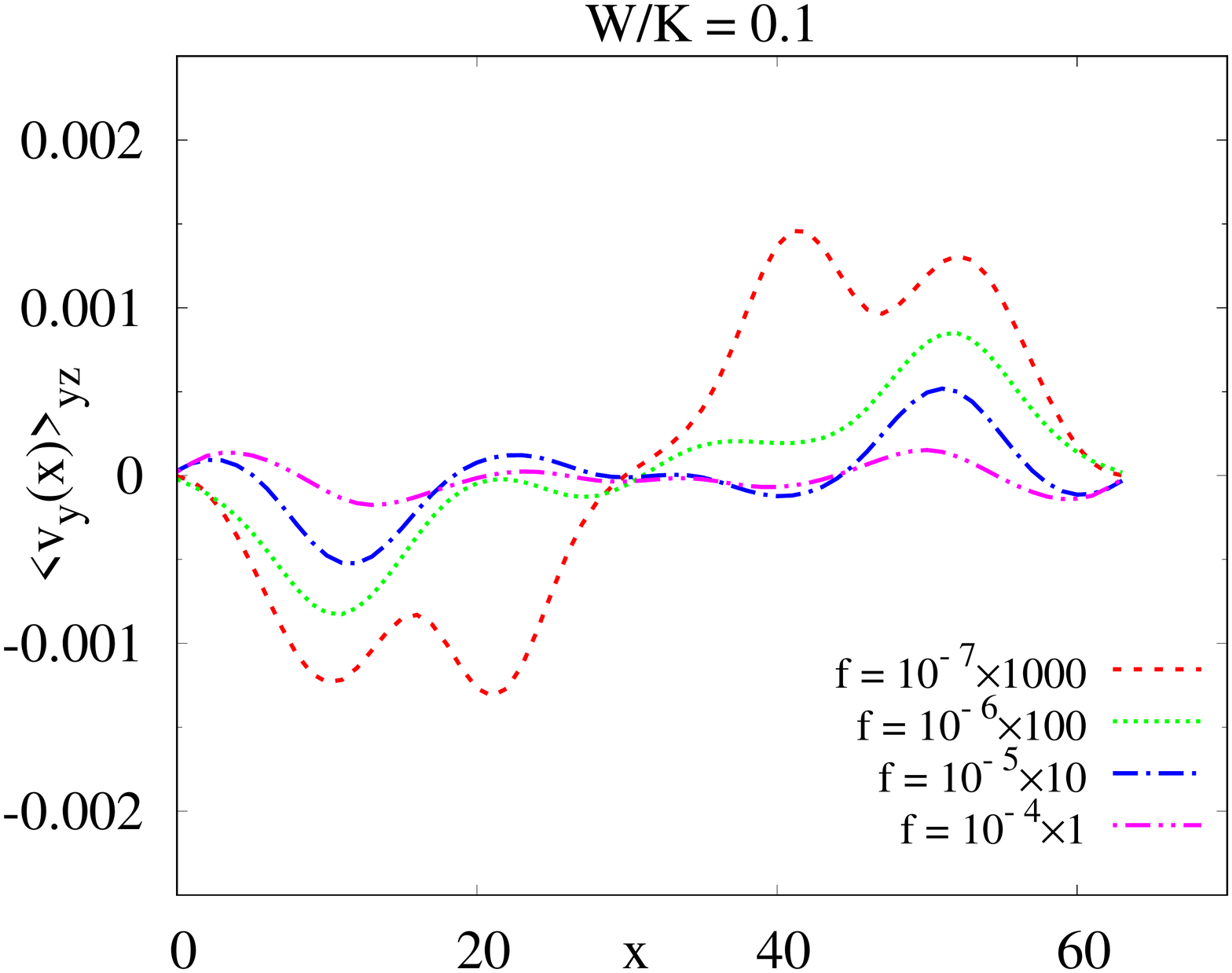}
\includegraphics[width=0.495\textwidth]{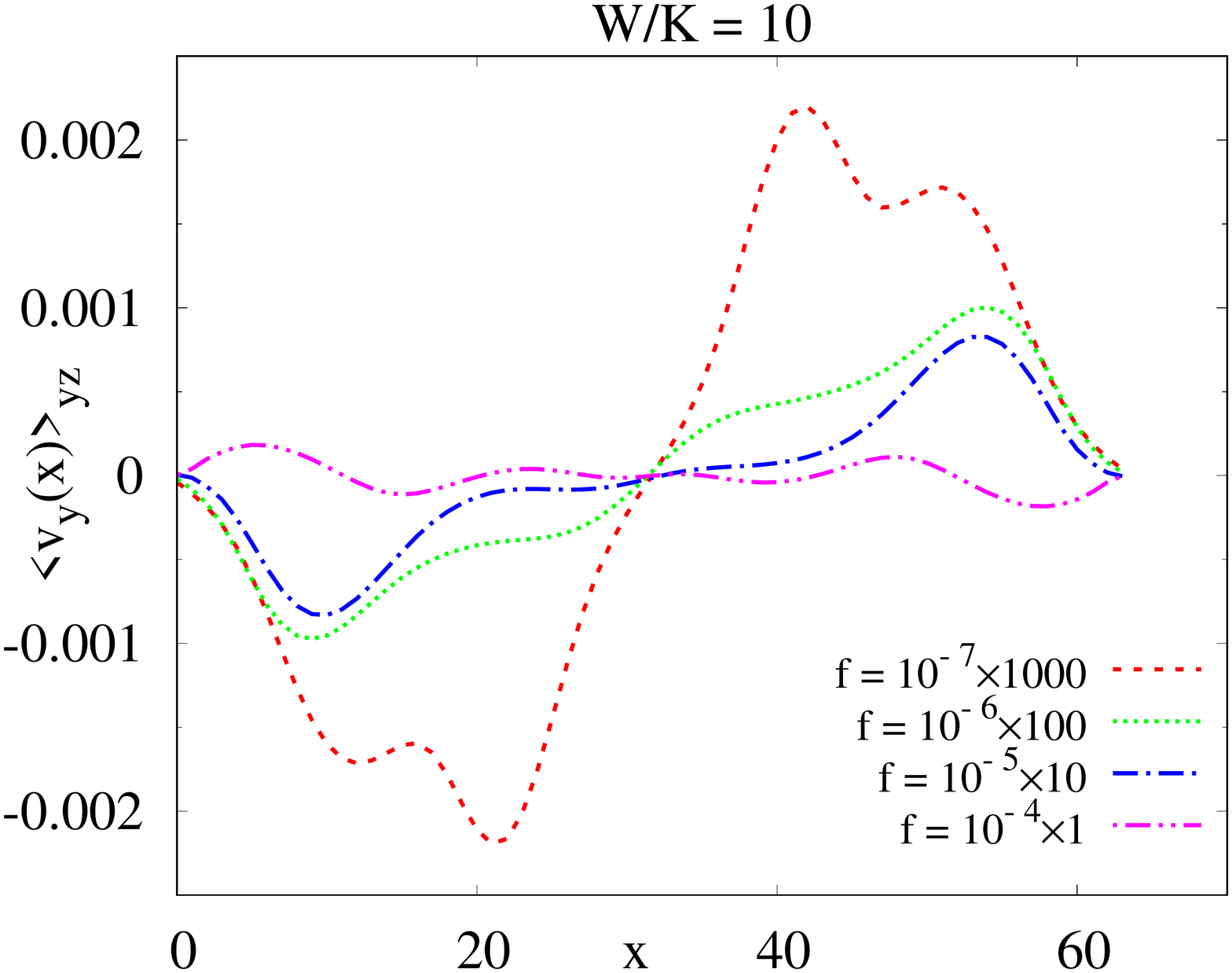}\\
\caption{\label{BP2_plan_velocity} Flow velocity profiles at different pressure gradients $f$ for weak (left column) and strong (right column) planar degenerate anchoring. The top row shows y- and z-direction averaged snapshots of the z-component of the flow velocity $\langle v_z(x)\rangle_{yz}$ across the gap along the x-direction. The bottom row gives y- and z-direction averaged snapshots of the y-component of the flow velocity $\langle v_y(x)\rangle_{yz}$. Note that the data has been scaled to make it comparable to the data obtained for the largest applied pressure gradient.}
\end{center}
\end{figure*}
 
Fig. \ref{BP2_plan_viscosity} shows the apparent viscosity versus applied pressure gradient for weak and strong planar degenerate anchoring. The curves for strong and weak anchoring are very close together, unlike for homeotropic anchoring where we observed stronger shear thinning for the larger anchoring strength (see Fig. \ref{BP2_norm_viscosity}). The apparent viscosity shows a plateau from $f=10^{-6}$ on and decreases gradually for larger pressure gradients in excess of $f=10^{-5}$. The overall trend, however, is qualitatively similar to the one observed for homeotropic anchoring. 
The reason why strong and weak anchoring lead to a very similar rheological response is readily understood from what was stated above. Planar degenerate anchoring conditions, whether strong or weak, promote the flow alignment of the director field near the wall, which we observe also at relatively modest pressure gradients.

Fig. \ref{BP2_plan_velocity} depicts the averaged velocity profiles across the gap for weak and strong planar degenerate anchoring.
For both anchoring strengths the y- and z-averaged velocity components along the flow direction $\langle v_z(x)\rangle_{yz}$ are now symmetric with respect to the centre of the channel, unlike for the case of weak homeotropic anchoring where an asymmetry was visible at low pressure gradients $f$ (see Fig.\ref{BP2_norm_velocity}). This is because the DTCs are now positioned symmetrically within the cell, as shown in Fig. \ref{BP2_plan_w4e-1}.
For the lowest pressure gradients $f=10^{-7}$ we observe again (as for homeotropic anchoring) a small residual modulation due to the stationary disclination network in the permeative mode of flow. This gives gradually way to a smoother profile as the pressure gradient increases and the DTC structure breaks up.
In line with what we observed in the plots of the apparent viscosity $\nu_{app}$, there is much more consistency in the y- and z-averaged profiles of the velocity component along the y-direction $\langle v_y(x)\rangle_{yz}$ between weak and strong planar degenerate anchoring as compared to the case of homeotropic anchoring. This applies particularly to the results at low pressure gradients $f=10^{-7}$ and $f=10^{-6}$. The profiles for the two larger pressure gradients $f=10^{-5}$ and $f=10^{-4}$ resemble very much those shown in Fig. \ref{BP2_norm_velocity}.

\section{Conclusion}\label{conclusions}

This study provides the first theoretical treatment of the pressure-driven, Poiseuille flow of cholesteric liquid crystals with a non-trivial two- or three-dimensional director field pattern \rev{subject to normal or planar degenerate anchoring conditions as they are commonly realised in experiments.} Indeed, previous computational studies assumed, for simplicity, either free or pinned boundary conditions, where a non-trivial director field pattern is fixed at the wall~\cite{Marenduzzo:2004,Dupuis:2005,Henrich:2012b}. \rev{Such boundary conditions were used as they are those which should least affect bulk ordering, and this is especially an issue if sample size is small (as the regions where boundaries affect ordering would extend to a significant part of the sample). In our case, we can use a more realistic boundary condition because the sample size is larger.}

While we limited our studies to Cholesteric Fingers of the first kind (CF1) and Blue Phase II (BPII), we expect that other BPs, or other Cholesteric Fingers, should lead to qualitatively similar behaviour. Our choice of phases was dictated by previous experience which suggest that these structures can be realised experimentally, and at the same time have a simpler response to bulk shear flow with respect to other cholesteric phases with 2D or 3D director field patterns ~\cite{Henrich:2013}. \rev{Because of the more complex order structure of the investigated cholesteric phases, we limited our study to the simplest possible microfluidic geometry, an infinitely long, one-dimensional channel. Nevertheless, and despite these limitations, we believe the results we present here are an important step towards real-world microfluidic applications where a more complicated geometry is coupled to the experimentally realistic boundary conditions we considered.} 

Our work shows that the \rev{different} anchoring conditions lead to large quantitatively differences with respect to \rev{previously reported results.} For instance, it was found that in the permeative mode the viscosity of a cholesteric liquid crystal with pinning boundary conditions was $30$-fold larger than at large flow ~\cite{Marenduzzo:2006b}. Here, we observe that for CF1, where we find the largest shear thinning, the apparent viscosity decreases only less than a factor of $2$ (Fig.\ref{CF1_viscosity}). Likewise, it was previously found that a system consisting of a single BPII cell with the director frozen at the wall had a viscosity which was over twice the Newtonian limit ~\cite{Dupuis:2005}. Here, in contrast the apparent viscosity of BPII is sizeably smaller. 
Notwithstanding these significant quantitative difference, strikingly, we find that qualitatively several phenomena are common to the case of pinned boundary conditions or bulk shear flow, and the homeotropic/planar degenerate anchoring we consider. First, in all cases there is permeative flow for very low pressure gradients ~\cite{Marenduzzo:2004}. Second, at larger pressure gradient we observe a continuous cycle of breakup and reformation of the disclination network of BPII, similarly to the case of bulk flow ~\cite{Henrich:2013}. Third, the bubble regime, corresponding essentially to floating DTCs, is reminiscent of the doubly-twisted texture observed for strongly forced cholesterics ~\cite{Marenduzzo:2004}. In this work, we find that this regime is ubiquitous, as floating DTCs are common in both CF1 and BPII, and we explain the mechanism leading to this pattern: it is generically observed when a strong enough shear at the boundary breaks the two- or three-dimensional cholesteric director field pattern, so that the region in the centre does not feel the constraining torque from the wall.

We characterise the sequence of dynamical, flow-induced regimes, which we predict CF1 and BPII structures should display when subjected to a pressure driven flow in a microfluidic geometry with controlled anchoring conditions. Such experiments are now feasible, with techniques similar to those reported in a recent experimental work ~\cite{experimentBPrheology}.
For CF1s we only considered \rev{strong and weak} homeotropic anchoring conditions, as these are crucial for obtaining a stable CF1 phase without flow. We identified four different dynamic regimes. At very low pressure gradients we observe a permeative regime where the finger pattern remains static whilst having a finite mass flow through the channel. At larger pressure gradients the finger pattern drifts downstream but remains 'attached' to one side of the channel breaking the symmetry with respect to its centre. The reasons for this emerging asymmetry are the backflow mechanism together with the fact that CF1s have a point symmetry with respect to the finger axis. At even larger pressure gradients the CF1s 'detach' from the second wall and form symmetric bubble-like conformations at the centre of the channel. This change in the local order structure entails a pronounced decrease in the apparent viscosity of the flowing CF1. At the largest applied pressure gradient the CF1 transforms into a Grandjean texture where the cholesteric helical axis is oriented perpendicular to the flow direction and the walls.

BPII has been simulated with both \rev{homeotropic and planar degenerate} anchoring conditions and at different anchoring strengths in a parameter region where it is thermodynamically stable in bulk. Following the permeative regime at the lowest pressure gradient (associated with strongly non-parabolic velocity profiles), we find for slightly larger forcing a regime in which the BPII disclination network drifts downstream. BPII shows the previously mentioned periodic breakup and reconnection of the disclination network ~\cite{Henrich:2013}. In the drifting regime, the network also slowly moves along the neutral direction, i.e., perpendicular to both the flow and flow gradient directions. We observe a gradual breakup of the double twist cylinders (DTCs) with increasing pressure gradient, starting at the walls where the velocity gradients are largest and then continuing from both sides towards the centre of the channel. At the largest pressure gradients only one floating row of DTCs remains at the centre of the channel whereas everywhere else the liquid crystal adopts a Grandjean-like texture with the cholesteric helical axis perpendicular to the flow direction and the walls. While we do not observe the Grandjean or nematic texture in the BPII case, we would expect that these may be observed for yet larger forcing, which cannot be reached in our LB simulations.

Finally, the apparent viscosity declines in all cases. This is expected as all systems we consider should shear thin, as found both in other simulations and experimentally ~\cite{experimentBPrheology}. The decrease of the apparent viscosity with increasing pressure gradient is more gradual for BPII than for CF1, which reflects the gradually dissolving DTC structure in the former. It is also interesting to note that, again for the BPII case and planar degenerate anchoring, the apparent viscosity is practically independent of the anchoring strength. This is however not the case for normal anchoring. Our interpretation is that this is due to the fact that planar anchoring is more compatible with flow alignment of the director field close to the walls. 

\section{Acknowledgments}

We acknowledge support from the EPSRC Programme Grant 'Design Principles for New Soft Materials' (EP/J007404/1). OH acknowledges support from the EPSRC Research Software Engineer Fellowship Scheme (EP/N019180/1). We acknowledge very useful discussion with Dr Anupam Sengupta. This work used the ARCHER UK National Supercomputing Service.

\footnotesize{
\bibliography{permeative} 
\bibliographystyle{rsc} 
}

\end{document}